\documentclass[aps,twocolumn,bibnotes,nofootinbib,superscriptaddress]{revtex4}

\usepackage{graphicx}
\usepackage{amssymb}
\usepackage{amsmath}
\usepackage{amsfonts}
\usepackage{color}
\usepackage{hyperref}
\usepackage{bm}

\DeclareMathAccent{\ring}{\mathalpha}{operators}{"17}
\providecommand{\st}[1]{_{\text{#1}}}

\providecommand{\ut}[1]{^{\text{#1}}}

\def\pd{\partial}
\def\tr{\mathrm{Tr}}

\def\drm{\mathrm{d}}

\def\qv{\bv{q}}
\def\uv{\bv{u}}

\def\Fv{\bv{F}}
\def\fv{\bv{f}}

\def\pv{\bv{p}}
\def\rv{\bv{r}}

\def\b0{\bv{0}}

\def\in{\st{in}}

\def\ra{\rightarrow}

\newcommand{\mods}{\ensuremath{\kappa_{\text{S}}}}
\newcommand{\modb}{\ensuremath{\kappa_{\text{B}}}}
\newcommand{\moda}{\ensuremath{\kappa_\alpha}}
\newcommand{\modA}{\ensuremath{\kappa_{\text{A}}}}
\newcommand{\modV}{\ensuremath{\kappa_{\text{V}}}}

\newcommand{\TBTT}{tumbling-to-tank-treading transition}

\newcommand{\bitem}{\begin{itemize}}
\newcommand{\eitem}{\end{itemize}}
\newcommand{\benum}{\begin{enumerate}}
\newcommand{\eenum}{\end{enumerate}}
\newcommand{\bblock}[1]{\begin{block}{#1}}
\newcommand{\eblock}{\end{block}}
\newcommand{\bmini}[1]{\begin{minipage}{#1}}
\newcommand{\emini}{\end{minipage}}
\newcommand{\btab}[1]{\begin{tabular}{#1}}
\newcommand{\etab}{\end{tabular}}
\newcommand{\btabn}[1]{\begin{tabular}{#1}}
\newcommand{\etabn}{\end{tabular}}
\newcommand{\beq}{\begin{equation}}
\newcommand{\eeq}{\end{equation}}
\newcommand{\beqn}{\begin{equation*}}
\newcommand{\eeqn}{\end{equation*}}
\newcommand{\bmult}{\begin{multline}}
\newcommand{\emult}{\end{multline}}
\newcommand{\bsplit}{\begin{split}}
\newcommand{\esplit}{\end{split}}

\newcommand{\bv}[1]{\mathbf{#1}}
\newcommand{\Ca}{\ensuremath{\text{Ca}}}

\newcommand{\Caeff}{\ensuremath{\text{Ca}^*}}

\begin{document}

 \title{Rheology of dense suspensions of elastic capsules: normal stresses, yield stress, jamming and confinement effects}
 \author{Markus Gross}
 \email{markus.gross@rub.de}
 \affiliation{Interdisciplinary Centre for Advanced Materials Simulation (ICAMS), Ruhr-Universit\"at Bochum, Universit\"atsstr. 150, 44780 Bochum, Germany}
  \author{Timm Kr\"uger}
\affiliation{Institute for Materials and Processes, School of Engineering, University of Edinburgh, King's Buildings, Mayfield Road, Edinburgh EH9 3JL, Scotland, UK}
 \affiliation{Centre for Computational Science, University College London, 20 Gordon Street, London WC1H 0AJ, UK}
\author{Fathollah Varnik}
 \affiliation{Interdisciplinary Centre for Advanced Materials Simulation (ICAMS), Ruhr-Universit\"at Bochum, Universit\"atsstr. 150, 44780 Bochum, Germany}
 \affiliation{Max-Planck Institut f\"ur Eisenforschung, Max-Planck Str.~1, 40237 D\"usseldorf, Germany}

\begin{abstract}
We study the shearing rheology of dense suspensions of elastic capsules, taking aggregation-free red blood cells as a physiologically relevant example. Particles are non-Brownian and interact only via hydrodynamics and short-range repulsive forces. An analysis of the different stress mechanisms in the suspension shows that the viscosity is governed by the shear elasticity of the capsules, whereas the repulsive forces are subdominant. Evidence for a dynamic yield stress above a critical volume fraction is provided and related to the elastic properties of the capsules. The shear stress is found to follow a critical jamming scenario and is rather insensitive to the tumbling-to-tank-treading transition. The particle pressure and normal stress differences display some sensitivity to the dynamical state of the cells and exhibit a characteristic scaling, following the behavior of a single particle, in the tank-treading regime. The behavior of the viscosity in the fluid phase is rationalized in terms of effective medium models. Furthermore, the role of confinement effects, which increase the overall magnitude and enhance the shear-thinning of the viscosity, is discussed.
\end{abstract}


\maketitle

\section{Introduction}
Non-Brownian suspensions of soft particles are not only relevant from a technological or biological perspective, but they are also a paramount example of soft glassy materials and understanding of their flow behavior has gained much attention recently \cite{stickel_fluid_2005, coussot_rheophysics_2007, schall_shear_2010}.
As a consequence of the deformable nature of the particles, the suspension attains viscoelastic properties and the effective viscosity shows a pronounced dependence on shear rate \cite{larson_structure_1999, pal_rheology_2007}.

Blood is a special, physiologically relevant example of an athermal soft-particle suspension, consisting mainly of red blood cells (RBCs) in a liquid medium. 
Both the overall rheological behavior of blood (see, e.g., \cite{popel_microcirculation_2005, yilmaz_critical_2008} and references therein) as well as the hydrodynamics of isolated vesicles and capsules (see, e.g., \cite{pozrikidis_computational_2010, barthes-biesel_modeling_2011, misbah_vesicles_2012}) have been studied in quite detail.
However, connecting the dense and dilute regime is challenging owing to the complex shape, dynamics and interactions of the particles.
The coupling between flow and particle deformation complicates theoretical approaches to the rheology of soft-particle suspensions or emulsions \cite{yaron_viscous_1972, choi_rheological_1975, brennen_concentrated_1975, pal_rheology_2011}, making simulations often an indispensable tool \cite{eggleton_large_1998, zhang_immersed_2007, dupin_modeling_2007, bagchi_mesoscale_2007, doddi_three-dimensional_2009, macmeccan_simulating_2009, clausen_capsule_2010, clausen_rheology_2011, reasor_rheological_2013}.

Due to their shear elasticity and non-spherical shape, RBCs usually tumble at low and tank-tread at high shear rates \cite{schmid-schonbein_fluid_1969, abkarian_swinging_2007, skotheim_red_2007}, and different views exist as to how these behaviors affect the macroscopic rheological properties of blood \cite{schmid-schonbein_fluid_1969, pfafferott_red_1985, chien_red_1987, forsyth_multiscale_2011}. 
Furthermore, the physical origin of a yield stress \cite{chien_effects_1966, picart_human_1998}, contributing to the strong increase of viscosity at high volume fractions, has remained somewhat unclear in the case of non-aggregating RBC suspensions
Also, normal stresses, while being studied for a long time for other types of viscoelastic fluids \cite{larson_structure_1999}, are largely unexplored in the case of suspensions of deformable particles, such as blood \cite{reasor_rheological_2013}. 
Normal stresses induce cross-streamline migration of particles \cite{nott_pressure-driven_1994, morris_curvilinear_1999, yurkovetsky_particle_2008, deboeuf_particle_2009} and can lead to inhomogeneous concentration profiles in non-axisymmetric flows \cite{ramachandran_influence_2008,malkin_rheology:_2006}.

In this work, we shed a light on the above issues via analytical models as well as numerical simulations of dense suspensions of non-aggregating RBCs.
Most of our insights and conclusions, however, do not rely on the particular shape of RBCs and should thus be applicable to other types of athermal capsule suspensions as well. 
We simulate suspensions of non-aggregating RBCs in wall-driven shear flow by coupling, via the immersed boundary method \cite{peskin_immersed_2002}, a finite element model for the capsule mechanics with a lattice Boltzmann model for the hydrodynamics \cite{kruger_efficient_2011, kruger_crossover_2013}. 
The mechanical properties of an RBC are described by Skalak's \cite{skalak_strain_1973} and Helfrich's \cite{helfrich_elastic_1973} constitutive laws for the shear and bending energies, together with additional constraints to ensure conservation of membrane area and volume.
Short-range repulsive interactions are included to improve numerical stability and are found to have a negligible influence on the rheology.
Our simulations cover three decades in reduced shear rate (capillary number) and volume fractions $\phi$ between 12 and 90\%, which significantly exceed comparable previous simulation works \cite{macmeccan_simulating_2009, fedosov_predicting_2011, clausen_rheology_2011, reasor_rheological_2013} and thus allow us to unveil crucial scaling laws governing the suspension rheology. 
We provide evidence for the existence of a yield stress above a critical volume fraction in the present, aggregation-free model, and rationalize it in terms of Hertzian contact elasticity. 
For intermediate shear rates, the viscosity is strongly shear-thinning, following an effective power-law.
Our results indicate that cell elasticity and the distance to the jamming point are the most dominant factors determining the rheology of suspensions of aggregation-free RBCs or similar types of capsules.
The rotational dynamics of the cells, in contrast, is not found to have a significant impact on the suspension rheology.
An exception is the particle pressure and first normal stress difference, which are sensitive to the \TBTT\ and show a scaling behavior in the tank-treading regime at high capillary numbers.
To gain a principal understanding of the viscosity in the dense regime, different effective medium models are investigated and the role of confinement effects is analyzed.
We find that wall-induced confinement of a capsule in a shear flow enhances the shear-thinning of the single-particle viscosity and argue that similar effects should also play a role in a dense suspension, where an effective confinement is provided by the neighboring particles.
Indeed, equipping an effective medium model with phenomenological confinement corrections leads to a remarkably accurate prediction of the simulated viscosity up to concentrations of around 40\%. Beyond that value, important features such as the power-law exponent describing the shear-thinning of the viscosity are still correctly captured.

\section{Methodology and Simulations}

\subsection{Simulation model}
The mechanical properties of the capsules in our simulations are modeled in terms of an energy functional of the form
\begin{equation}
 E = E_S + E_B + E_A + E_V\,,
\label{eq_energy_tot}
\end{equation}
where the individual terms describe energy penalties against shear, bending, area and volume changes, respectively.
For the shear energy, we employ Skalak's model \cite{skalak_strain_1973},
\begin{equation}
 E_S = \oint \drm A \left[ \frac{\kappa_S}{12} (I_1^2 + 2I_1 -2I_2) + \frac{\kappa_\alpha}{12}I_2^2 \right]\,,
\label{eq_shear_energy}
\end{equation}
with $\mods$ and $\moda$ being the shear and area modulus and $I_{1,2}$ the in-plane strain invariants, which are related to the eigenvalues of the local membrane deformation tensor (see \cite{kruger_efficient_2011} for more details).
The bending energy is described by Helfrich's model \cite{helfrich_elastic_1973},
\begin{equation}
 E_B = \frac{\kappa_B}{2} \oint \drm  A\left( H - H^{(0)}\right)^2 \,,
\label{eq_bending_energy}
\end{equation}
where $\modb$ is the bending modulus, $H$ and $H^{(0)}$ are the mean and spontaneous curvatures.
While the resistance against surface area changes embodied in eq.~\eqref{eq_shear_energy} is due to the cytoskeleton, a much larger energy penalty arises from the bilayer, which can be described by a surface energy of the form \cite{evans_mechanics_1980, seifert_configurations_1997},
\begin{equation}
  E_A = \frac{\kappa_A}{2} \frac{(A-A^{(0)})^2}{A^{(0)}}\,,
\label{eq_area_energy}
\end{equation}
with $\modA$ being the surface modulus.
Finally, the conservation of cell volume is ensured by means of a volume energy,
\begin{equation}
  E_V = \frac{\kappa_V}{2} \frac{(V-V^{(0)})^2}{V^{(0)}}\,,
\label{eq_volume_energy}
\end{equation}
with $\modV$ being the volume modulus. In the above, $A$, $V$ and $A^{(0)}$, $V^{(0)}$ are the instantaneous and equilibrium values of the membrane area and volume, respectively.
We note that $E_V$ is introduced primarily due to numerical reasons related to the immersed boundary method \cite{peskin_improved_1993}, although its shape can be motived based on considerations of the osmotic pressure of the RBC \cite{seifert_configurations_1997}.
For further information on the physical origin of the above mechanical laws, we refer to the literature \cite{evans_mechanics_1980,svetina_cooperative_2004,gompper_soft_2008}.

In our simulations, a capsule is represented by a moving (Lagrangian) mesh obtained from triangulation of the biconcave surface of an RBC. 
From eq.~\eqref{eq_energy_tot}, the elastic forces acting on each membrane node located at position $\rv_i$ are obtained via
\beq
\fv_i\ut{el} = -\frac{\pd E(\{\rv_i\})}{\pd \rv_i}\,.
\label{eq_mem_force_el}
\eeq
To improve numerical stability and avoid potential particle overlap, we additionally employ a repulsive force between any two nodes of two cells in proximity. This force respects conservation of linear and angular momentum, is zero for node-to-node distances larger than one lattice constant and increases as $1/r^2$ for smaller distances (cf.~\cite{feng_immersed_2004}):
\beq \fv\ut{int}_{ij} = \left\{\begin{aligned}-\kappa\st{int}(d_{ij}^{-2} - 1) \frac{\bv{d}_{ij}}{d_{ij}} \qquad & \text{for}\qquad d_{ij}<1,\\ 0 \qquad & \text{for}\qquad d_{ij}\geq 1\,. \end{aligned}\right.
\label{eq_mem_force_int}
\eeq
Here, $\bv{d}_{ij}$ is the distance between the two nodes $i$ and $j$, $\kappa\st{int}$ is a constant parameter and $\fv_{ji}\ut{int} = -\fv_{ij}\ut{int}$.

The suspending fluid is governed by the Navier-Stokes equations,
\begin{align} 
\pd_t \rho &= -\nabla\cdot(\rho\uv)\,, \\
\begin{split}
\pd_t(\rho\uv) + \nabla\cdot(\rho\uv\uv) &= -\nabla P + \eta_0\nabla^2\uv  \\&\quad +(\zeta_0+\eta_0/3)\nabla\nabla\cdot \uv + \Fv\,,
\end{split}
\end{align}
where $\rho$ and $\uv$ are the density and velocity, $\eta_0$, $\zeta_0$ are bare shear and bulk viscosities and $\Fv$ is an external force density (see below). The Navier-Stokes equations are solved using a standard ideal-gas type lattice Boltzmann algorithm \cite{kruger_efficient_2011}. The pressure $P$ is given by $P=\rho/3$ for the present lattice Boltzmann model \cite{succi_lattice_2001}. Although the present method admits in principle for inertial effects, we select parameters such that the Reynolds and Mach number (the latter being an indicator of the compressibility of the fluid) remain small.

The coupling between fluid and solid is realized via the immersed boundary method \cite{peskin_immersed_2002}, in which the total membrane force $\fv\ut{tot}$ is ``spread'' to the Eulerian fluid grid according to (a suitably discretized form of)
\beq \Fv(\qv) = \int \drm s\, \drm v\, \fv\ut{tot}(s,v)\delta(\qv-\rv(s,v))\,.
\eeq
Here, $\qv$ denotes the position of a fluid node, $\rv$ denotes a location on the surface of the membrane parametrized by $s$ and $v$. The total membrane force is given by the sum of the elastic [eq.~\eqref{eq_mem_force_el}] and interaction forces [eq.~\eqref{eq_mem_force_int}], 
\beq \fv\ut{tot} =  \fv\ut{el} + \fv\ut{int}\,.
\label{eq_force_tot}
\eeq
Conversely, the membrane nodes are moving with the local flow velocity interpolated at each membrane node:
\beq \dot\rv(s,v) = \int \drm^3q\, \uv(\qv)\delta(\qv-\rv(s,v))\,.
\eeq
We generally employ a two-point stencil for the evaluation of the discrete version of the delta function in the above equations. Further details can be found in \cite{kruger_efficient_2011}.

\subsection{Stress evaluation}
In the lattice Boltzmann method, several approaches are available to compute global and local stresses. 
In the presence of solid walls, the simplest method to obtain the system-averaged shear stress is to consider the momentum exchange $\Delta \pv$ between fluid and wall,
\beq \frac{\Delta \pv}{\Delta t}= \boldsymbol{\sigma}^w \cdot \Delta \bv{A}\,.
\label{eq_wallstress}\eeq
Here, $\Delta \bv{A}$ is a normal vector of an area element (with area $\Delta A$) at the wall and $\boldsymbol\sigma^w$ is the wall stress. 
Assuming that particles experience no direct contact with the wall, the momentum transfer $\Delta\pv$ can be simply computed form the lattice Boltzmann populations subject to the bounce back rule at the wall, see \cite{ladd_numerical_1994, ladd_numerical_1994-1, ladd_lattice-boltzmann_2001}. 
In this case, the total shear stress of the suspension is given by $\sigma^w_{xz} = \pm \Delta p_x \Delta t \Delta A$, assuming wall normals pointing in $z$-direction and shear flow in $x$-direction.
If particles can interact directly with the wall, the corresponding interaction forces have to be added to $\Delta \pv$ to ensure momentum conservation. This applies, for instance, when particles are glued to the wall in order to mimic roughness.

In cases where the full tensorial information on the stress is needed, one might resort to the stresslet approach due to Batchelor \cite{batchelor_stress_1970}.
Here, the bulk stress of an overall force-free suspension is split into a contribution of the pressure in the fluid volume, the bare fluid stress (i.e., the stress in the absence of particles) and the stress due to the particles:
\beq \boldsymbol\sigma = - P \bv{I} + 2\eta_0 \bv{E} + \boldsymbol{\sigma}^P\,,
\label{eq_stresslet}\eeq
with $P$ and $\bv{E}$ being the volume-averaged scalar pressure and strain tensor.
The average particle stress $\boldsymbol{\sigma}^P$ consists, in the present case, of a contribution from elastic membrane and particle interaction forces,
\beq \boldsymbol{\sigma}^P = \boldsymbol{\sigma}^{P,\text{el}} + \boldsymbol{\sigma}^{P,\text{int}}\,.
\label{eq_partStresslet_tot}
\eeq
For liquid filled membranes with identical inner and outer viscosities, we have \cite{pozrikidis_finite_1995, bagchi_rheology_2010} 
\beq \sigma_{\alpha\beta}^{P,\text{el}} = -\frac{1}{V} \sum_k \sum_{i_k} \, f_{i_k,\alpha}\ut{el} r_{i_k,\beta}\,,
\label{eq_partStresslet_el}
\eeq
where $V$ is the total fluid volume, the first sum runs over all particles and the second over all membrane nodes $i_k$ of particle $k$. Here, $\bv{f}\ut{el}$ is the force exerted by the membrane on the fluid [eq.~\eqref{eq_mem_force_el}], which also equals the stress jump across the membrane surface.
The contribution from the interaction forces [eq.~\eqref{eq_mem_force_int}] to the stresslet of each participating particle is given by 
\beq \sigma_{\alpha\beta}^{P,\text{int}} = -\frac{1}{2V} \sum_k \sum_{i_k}\sum_{j_n} \, f_{i_k j_n,\alpha}\ut{int} d_{i_k  j_n, \beta}\,,
\label{eq_partStresslet_int}
\eeq
where $i_k$ and $j_n$ run over all nodes of the two interacting capsules $k$, $n$, and $\bv{d}_{i_k  j_n}$ is the distance vector connecting two such nodes.
We generally find that the stress computed from the momentum transport at the walls [eq.~\eqref{eq_wallstress}] is identical to the one obtained from the stresslet [eq.~\eqref{eq_stresslet}]. 

The effective suspension viscosity is defined in terms of the effective shear stress $\sigma_{xz}$ [eq.~\eqref{eq_stresslet}] as
\beq \eta =\frac{\sigma_{xz}}{\dot\gamma}\,.
\label{eq_visc}
\eeq
The viscosity of the particle phase alone is given by
$\eta^p \equiv \eta - \eta_0= \sigma^P_{xz}/\dot\gamma$.
The quantity $\eta^P/\eta_0 \phi$ is also called \emph{intrinsic} viscosity of the particle phase.
Besides the shear stress, the diagonal components of the stress tensor are of interest as well, which are typically studied in terms of particle pressure $\Pi$ and first and second normal stress differences, $N_1$, $N_2$:
\beq
\Pi = -\frac{1}{3}\tr \boldsymbol\sigma^P\,,\qquad N_1 = \sigma^P_{xx} - \sigma^P_{zz}\,,\qquad N_2 = \sigma^P_{zz} - \sigma^P_{yy}\,.
\label{eq_normstress}\eeq

\subsection{Simulation setup}
\begin{figure}[t]\centering
	\includegraphics[width=0.8\linewidth]{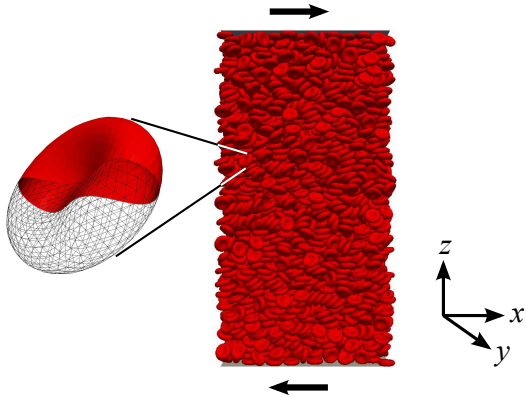}
	\caption{Sketch of the simulation setup of a typical system (here, $\phi=66\%$). Walls move with constant velocity in $\pm x$-direction. The magnification on the left shows the triangulated surface mesh of an RBC.}
    \label{fig_setup}
\end{figure}

\begin{table*}[t]
 \begin{tabular}{c|ccccccc}
\hline\hline
bare capillary number (Ca)     & 0.081 & 0.056 & 0.028 & 0.0028 & $2.8\times 10^{-4}$ & $8.4\times 10^{-5}$ & $4.2\times 10^{-5}$ \\
shear rate ($\dot\gamma/10^{-4}$) & 1.6   & 1.1   & 0.56  & 0.056  & 0.19                 & 0.11 & 0.056 \\
shear modulus ($\mods$)   & 0.003 & 0.003 & 0.003 & 0.003  & 0.1                  & 0.2 & 0.2 \\
\hline\hline
 \end{tabular}
\caption{Shear rate and elastic modulus (in lattice units) corresponding to the different bare capillary numbers used in this work. The bending modulus is taken as $\modb=\mods/5$ and the area and surface deviation moduli are fixed at $\moda = \modA = \modV = 1$.}
\label{tab_param}
\end{table*}

Three-dimensional simulations of RBCs in wall-driven shear flow are performed in a box of size $L_x\times L_y\times L_z = 180\times 180\times 360$ lattice units (l.u.).
The RBCs assume, in equilibrium, a biconcave disk shape of large semiaxis $r=9$ l.u.\ and are filled with a fluid having the same viscosity $\eta_0$ as the surrounding medium.
The mesh of the capsules consists of 1620 triangular facets and 812 nodes.
The total number of cells ranges between 1000 and 7700.
A layer of cells is glued to the wall in order to avoid slip effects. The corresponding cells are moving with the wall by means of a force that is proportional to the difference between the expected and actual particle position.
The shear rate profile is homogeneous on average and we do not observe long-time steady shear bands in our simulations \cite{mandal_heterogeneous_2012}. 
We have also checked, by studying the pair probability distribution (data not shown), that crystallization or lane formation \cite{ackerson_shear-induced_1988, ackerson_shear_1990, cheng_assembly_2012, zurita-gotor_layering_2012} is absent and the suspension remains in a disordered state. More details on the suspension microstructure will be published in a different article.

In order to minimize area and volume fluctuations of the cells, we generally set the area and volume deviation moduli to unity, $\moda = \modA = \modV = 1$ l.u., which is close to the upper limit of numerical stability.
To cover a large region of capillary numbers, we change both shear rate and shear modulus (see Tab.~\ref{tab_param}), keeping the ratio between bending and shear modulus fixed at $\modb/\mods = 2.47 \cdot 10^{-3}r^2$ ($\modb=\mods/5$ in l.u.). 
While it is known that the ratio of the shear modulus to the area dilation modulus has some effect on the shear thinning properties of a single capsule \cite{bagchi_rheology_2010}, we find that the suspension rheology is predominantly determined by the shear elasticity (cf.~Fig.~\ref{fig_stress_contrib} below) -- in particular, it is not significantly affected by varying the ratio $\moda/\mods$ or $\modA/\mods$. 
Thus, the capillary number defined by
\beq
 \Ca= \frac{\eta_0 \dot\gamma r}{\mods}\,,
\label{eq_Ca}
\eeq
is an appropriate dimensionless parameter to characterize the rheology of suspensions of elastic capsules. 
The capillary number can also be viewed as the ratio between the time scales of the shear relaxation of the capsule, $\tau\st{el}=\eta_0 r / \mods$, and external shear perturbation,
$\Ca = \tau\st{el} \dot{\gamma}$.
In addition to the \emph{bare} capillary number defined above, one might also define an \emph{effective} capillary number,
\beq
 \Caeff= \frac{\eta \dot\gamma r}{\mods}\,,
\label{eq_Caeff}
\eeq
based on the effective viscosity $\eta$ [eq.~\eqref{eq_visc}].
We have shown previously \cite{kruger_crossover_2013} that the rotational state (i.e., tumbling or tank-treading motion) of the capsules is sensitive to $\Caeff$. 
Thus, plotting a quantity versus $\Caeff$ can be a useful means to detect a possible influence of the cell rotation.

The volume fractions $\phi$ we report are effective quantities, corrected to account for the slightly increased hydrodynamic radius caused by force interpolation of the immersed boundary method \cite{kruger_efficient_2011}. 
We have determined the effective hydrodynamic radius by comparing the viscosity of a quasi-rigid spherical capsule with the classical Einstein prediction, $\eta=\eta_0(1+2.5\phi)$  \cite{einstein_new_1906}. 
We find that bare and effective volume fractions are related by $\phi\simeq (1.20\pm 0.05)\phi\st{bare}$ for the present resolution.
A possible alternative approach is based on the angular velocity of a quasi-rigid ellipsoidal capsule \cite{jeffery_motion_1922} and leads to similar results. 

\section{Results}
\subsection{Shear viscosity}

\begin{figure*}[t]\centering
	(a)\includegraphics[width=0.46\linewidth]{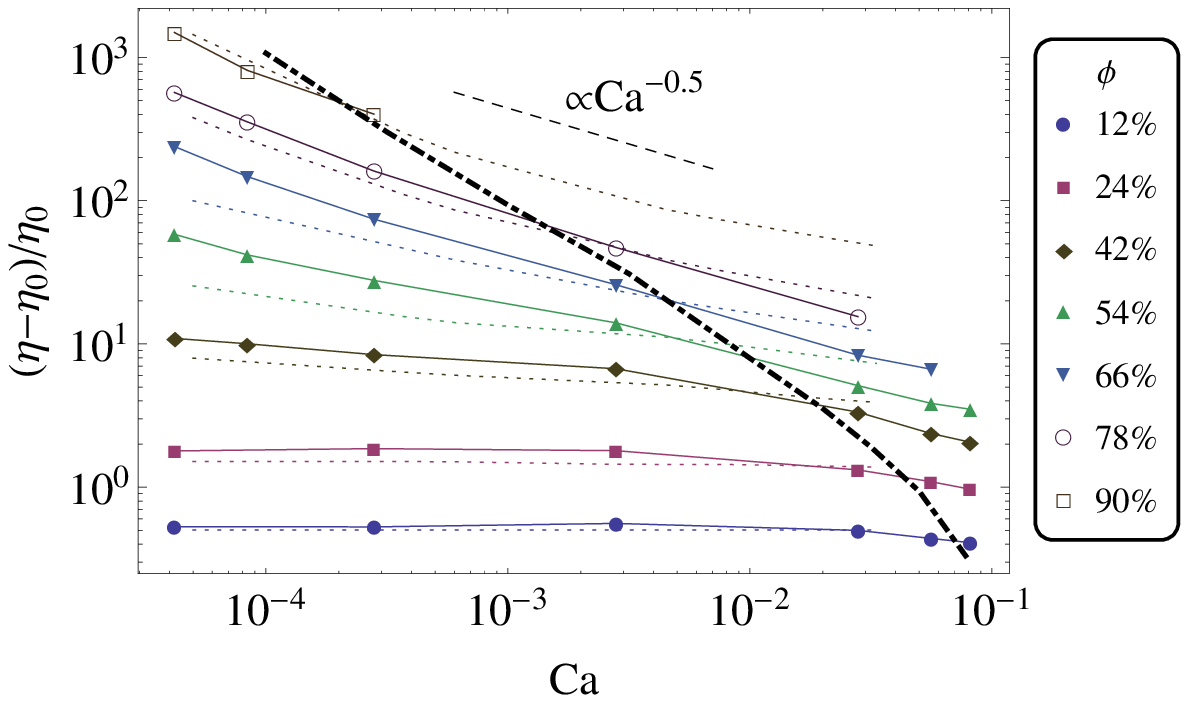}\quad
	(b)\includegraphics[width=0.46\linewidth]{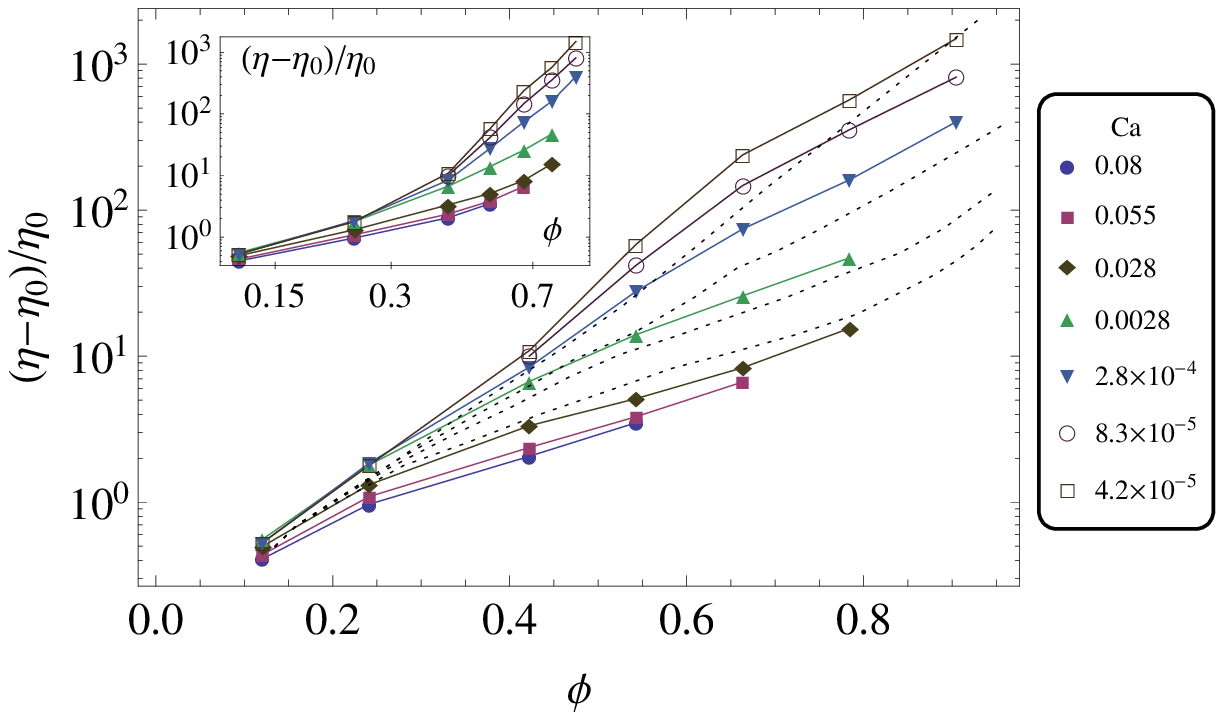}
	\caption{Effective suspension viscosity in dependence of (a) the bare capillary number and (b) the volume fraction as obtained from our simulations (symbols connected by solid lines). The inset in (b) shows the same data in double-logarithmic representation. Left to the dash-dotted curve in (a) cells perform tumbling motion, while they tank-tread right to it. The dotted curves represent experimental data extracted from \cite{chien_effects_1966}. In (a), these data have been interpolated over Ca for better comparison with the simulations. In (b), the original data set of \cite{chien_effects_1966} is shown, corresponding, from bottom to top, to $\Ca=0.05, 0.005, 5\times 10^{-4}, 5\times 10^{-5}$.}
    \label{fig_visc}
\end{figure*}

Fig.~\ref{fig_visc} shows the effective suspension viscosity $\eta$ (normalized to the viscosity $\eta_0$ of the background solvent) obtained from our simulations. 
Characteristic for suspensions of deformable particles \cite{larson_structure_1999, pal_rheology_2007}, the viscosity reveals a Newtonian plateau at low volume fractions and capillary numbers, which crosses over into a shear-thinning regime that becomes more pronounced with increasing volume fraction. 
We find that, at large capillary numbers and volume fractions, the viscosity follows an effective power-law $\eta\propto \Ca^{q}$, with $q\simeq -0.5$.
The behavior at still larger capillary numbers is not accessible in our simulations, but experiments on RBC suspensions have observed another Newtonian plateau at high shear rates \cite{schmid-schonbein_fluid_1969}.
The upward bending of the viscosity curves at large $\phi$ and small Ca is related to the presence of a yield stress (see sec.~\ref{sec_yieldstress}).
As Fig.~\ref{fig_visc}b shows, the viscosity grows approximately exponentially with $\phi$ and diverges at the maximum packing fraction $\phi=1$.
We remark that a power-law-like behavior over some range of $\phi$ can not be excluded, though (see inset to Fig.~\ref{fig_visc}b). 
A quantitative explanation of these observations in terms of effective medium models is provided in sec.~\ref{sec_effmed}.

The dotted curves in Fig.~\ref{fig_visc} represent experimental measurements of \cite{chien_effects_1966} performed on non-aggregating RBCs in Ringer solution. 
Experimental quantities were converted to suitable dimensionless numbers by making use of the typical physiological values of solvent viscosity $\eta_0=0.7\,\mathrm{mPa\,s}$, RBC radius $r=4\,\mu\mathrm{m}$ and shear modulus $\mods=5\,\mu\mathrm{N\,m^{-1}}$, which yield a relation between the capillary number and the experimentally reported shear rate, $\Ca= 9.6\cdot 10^{-4}\,\mathrm{s}\cdot \dot\gamma$.
Furthermore, since the viscosity values reported in \cite{chien_effects_1966} were given as a function of volume fraction for a limited number of different shear rates, we transformed them, for better comparison, in Fig.~\ref{fig_visc}a into the representation $\eta(\dot\gamma)$ via interpolation. 
Overall, we note a good agreement between the experimental measurements and our simulation results.
We remark that using the effective rather than the bare volume fraction is a crucial requirement for this comparison.
Part of the observed deviations can be attributed to the fact that, in the present simulations, fluids inside and outside of the cell have identical viscosities, whereas, in reality, the viscosity of the inner hemoglobin solution is about five times larger than the surrounding solvent. 
This is particularly relevant in the tank-treading regime at larger capillary numbers \cite{kruger_crossover_2013}, where the inner fluid is sheared and thus significantly contributes to dissipation.
Additionally, at larger $\Ca$, the membrane viscosity \cite{yazdani_influence_2013}, which is absent in the present model, is expected to become important.
These effects might also be responsible for the deviation of the experimental data from the effective power-law that describes the simulation results well.
We finally mention that, in principle, deviations could also be caused by a finite mesh resolution, especially at large volume fractions where particles are in close contact.
However, based on the good agreement with the experimental results, we expect these deviations to be of more quantitative nature and not to significantly affect the overall trends reported here.

\subsection{Effect of the \TBTT\ on the viscosity}
\begin{figure}[t]\centering
	\includegraphics[width=0.9\linewidth]{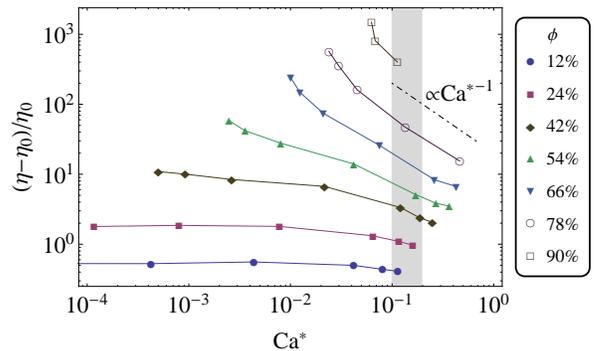}
	\caption{Suspension viscosity versus of effective capillary number, $\Caeff=\eta \dot\gamma r/\mods$, for different volume fractions. The shaded area marks the transition region from tumbling ($\Caeff\lesssim 0.1$) to tank-treading motion ($\Caeff\gtrsim 0.2$). In contrast to certain structural quantities, such as the nematic order parameter \cite{kruger_crossover_2013}, the effective viscosity shows no sharp transition at the \TBTT. Instead, a description in terms of a critical jamming scenario seems to be more appropriate (see text).}
    \label{fig_visc_Caeff}
\end{figure}

In a suspension, the crossover from tumbling to tank-treading motion happens at an effective capillary number $\Ca^*\equiv (\eta/\eta_0)\Ca\simeq 0.1$ (represented by the dashed curve in Fig~\ref{fig_visc}a) and is concomitant with a nematic ordering \cite{kruger_crossover_2013}. However, as Fig.~\ref{fig_visc} indicates, the transition between the two states has no obvious effect on the shear-thinning behavior of the viscosity. This is further emphasized by a plot of the suspension viscosity versus $\Caeff$ (Fig.~\ref{fig_visc_Caeff}), which does not reveal a sharp transition at the tumbling-to-tank-treading boundary. 
The absence of a strong influence of the \TBTT\ on the viscosity of a very dense suspension might not be surprising, though, as excluded volume effects are expected to dominate in that limit, hindering the free rotation of the cells. Indeed, it turns out that, at high densities, cells do not continuously tumble, but instead perform an intermittent flipping motion, where the typical time between flips can be of several ten inverse shear rates at high volume fractions \cite{kruger_crossover_2013}.

\begin{figure*}[t]\centering
	(a)\includegraphics[width=0.39\linewidth]{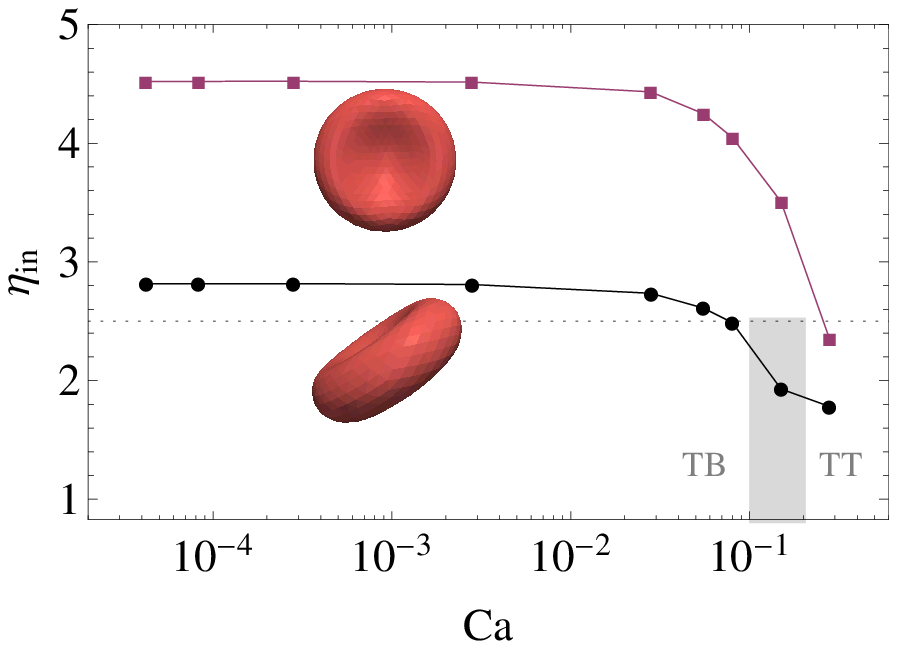}\quad
	(b)\includegraphics[width=0.42\linewidth]{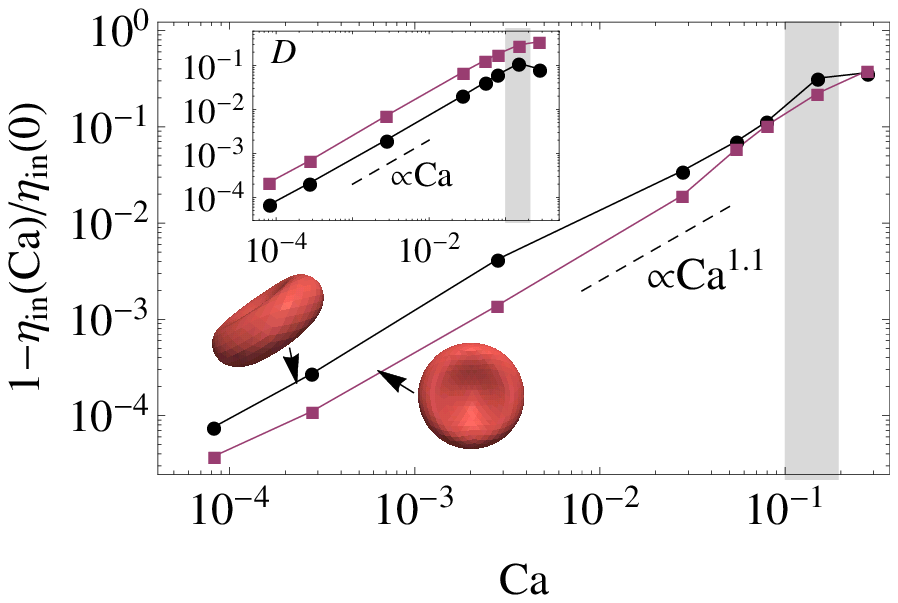}
	\caption{(a) Intrinsic viscosity $\eta\in\equiv (\eta-\eta_0)/\eta_0\phi$ of a single RBC in shear flow for two different orientations. The dotted line marks the intrinsic viscosity of a rigid sphere, $\eta\in=2.5$ \cite{einstein_new_1906}. (b) Relative deviation of the intrinsic viscosity $\eta\in$ from its limit for $\Ca\ra 0$ for the same data as in (a). The inset shows the corresponding Taylor deformation parameter $D$. The shaded area in (a,b) applies to the $\bullet$ and marks the transition regime from tumbling (TB) to tank-treading (TT) motion. }
    \label{fig_singlecell_visc}
\end{figure*}

Interestingly, however, even for a single RBC, we do not observe a clear sign of the \TBTT\ in the viscosity -- consistent with previous works on isolated ellipsoidal capsules \cite{bagchi_dynamic_2011, gao_shape_2012}. 
This is made directly evident by comparing (Fig.~\ref{fig_singlecell_visc}) the intrinsic viscosity $\eta\in \equiv (\eta-\eta_0)/\eta_0\phi$ of an RBC in shear flow for two different orientations, where, in one case, the cell tumbles at low Ca and tank-treads at high Ca, while in the other case, only the deformation changes with Ca. 
In both cases, the deviation of the intrinsic viscosity from its value for $\Ca\ra 0$, $\eta\in(\Ca)-\eta\in(0)$, behaves similarly and scales approximately linearly with capillary number (Fig.~\ref{fig_singlecell_visc}b). 
It is noteworthy that an approximately linear behavior of $\eta\in(\Ca)-\eta\in(0)$ is consistent with results of previous simulation works \cite{clausen_capsule_2010, bagchi_rheology_2010, bagchi_dynamic_2011}, but unexpected on the basis of a number of existing theoretical works on elastic capsules \cite{goddard_nonlinear_1967, roscoe_rheology_1967, barthes-biesel_constitutive_1981, navot_elastic_1998, gao_shape_2012}, which predict a scaling $\eta(0)-\eta(\Ca)\it\propto O(\text{Ca}^2)$ at leading order. The precise reason for this discrepancy is unclear at present, but might be related to the constitutive models employed and should be investigated in future works.
The cell deformation, as characterized by the Taylor deformation parameter, also depends linearly on Ca for $\Ca\lesssim 0.1$ (inset to Fig.~\ref{fig_singlecell_visc}b), in agreement with theoretical predictions \cite{barthes-biesel_motion_1980}.
The Taylor deformation parameter is defined as $D=(a-b)/(a+b)$, where $a$ and $b$ are the large and small major-axes of the equivalent inertia ellipsoid of the RBC (cf.~\cite{kruger_crossover_2013}).

\subsection{Stress mechanisms}
\label{stress_contrib}
\begin{figure*}[t]\centering
	(a)\includegraphics[width=0.38\linewidth]{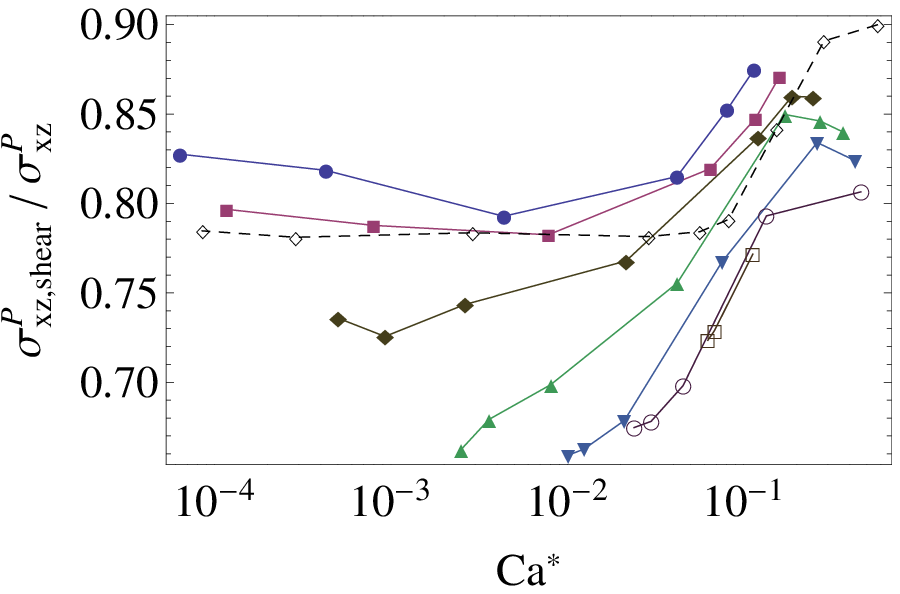}\qquad
	(b)\includegraphics[width=0.38\linewidth]{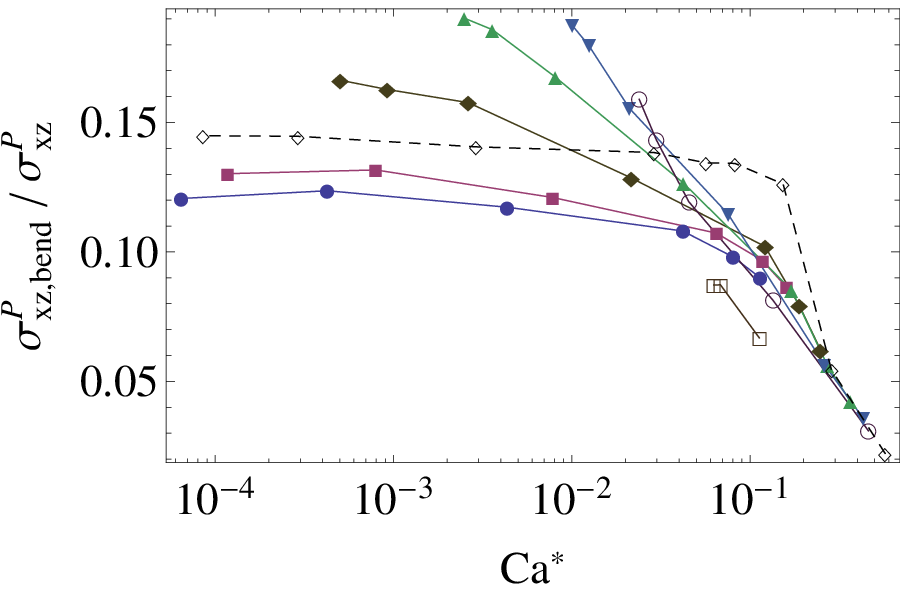}
	(c)\includegraphics[width=0.38\linewidth]{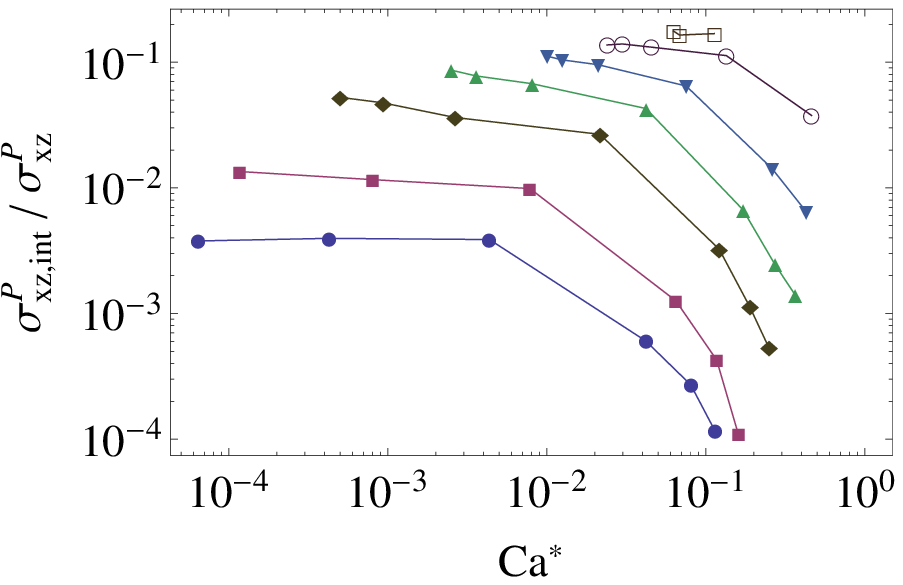}\qquad
	(d)\includegraphics[width=0.45\linewidth]{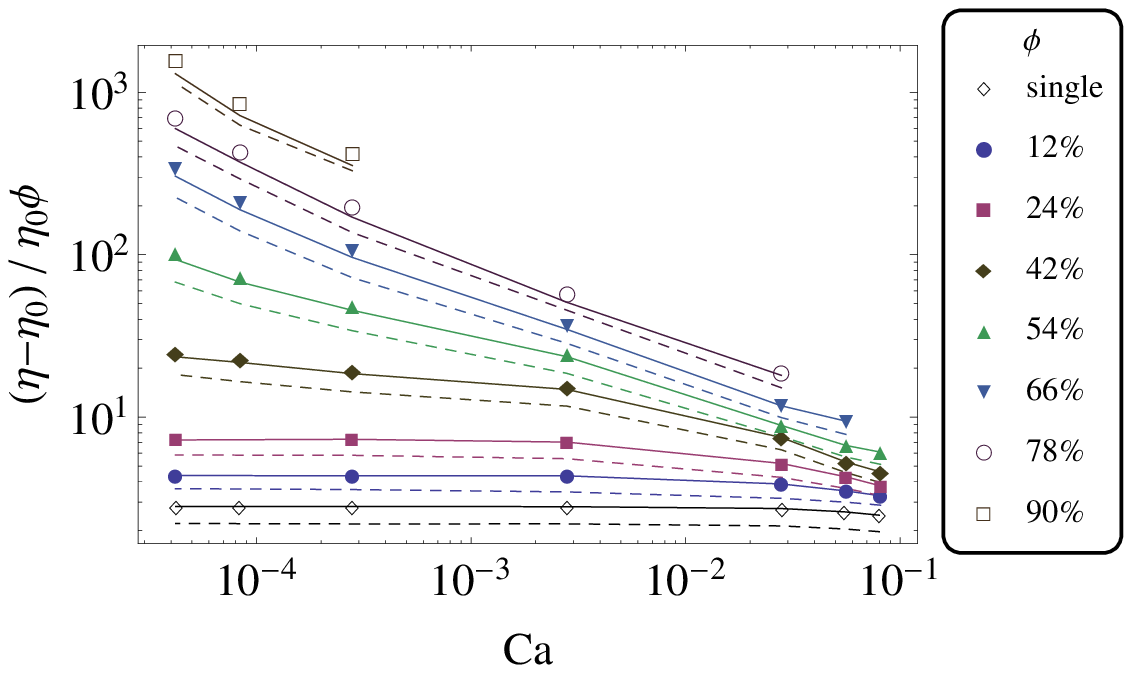}
	\caption{(a-c) Contributions to the particle shear stress [eq.~\eqref{eq_partStresslet_tot}] due to (a) shear, (b) bending, and (c) repulsive cell pair-interactions [eq.~\eqref{eq_partStresslet_int}]. All contributions are normalized to the total particle shear stress $ \sigma^P_{xz}$ and plotted vs.\ the effective capillary number $\Caeff$. The dashed line marks the behavior of a single RBC. (d) Comparison of the effective viscosity computed without interaction forces [$\eta^P=(\sigma^P_{xz}-\sigma^P_{xz,\text{int}})/\dot \gamma$, solid lines] and the viscosity based on the shear strain contribution alone ($\eta^P= \sigma^P_{xz,\text{shear}}/\dot \gamma$, dashed lines) to the total particle-phase viscosity ($\eta^P=\eta-\eta_0= \sigma^P_{xz}/\dot \gamma$, symbols). The legend in (d) applies to all panels.}
    \label{fig_stress_contrib}
\end{figure*}
Each of the forces arising in the model, eq.~\eqref{eq_force_tot}, contributes to the total particle stress in the simulation and can be accessed via the appropriate stresslets [see eqs.~\eqref{eq_partStresslet_el}, \eqref{eq_partStresslet_int}]. 
Fig.~\ref{fig_stress_contrib}a-c depict the magnitude of the shear components of the shear, bending, and interaction stresses relative to the total particle shear stress $ \sigma^P_{xz}$. 
Overall, the shear and bending stresses in the dense suspension follow the trend of a single RBC (dashed curves in Fig.~\ref{fig_stress_contrib}a-b).
The most dominant contribution to the stress is provided by the shear strain (Fig.~\ref{fig_stress_contrib}a), its magnitude being typically more than two thirds of the total particle stresslet.
The bending strain (Fig.~\ref{fig_stress_contrib}b) contributes most of the remaining part to the total stress in the tumbling regime ($\Ca^*\lesssim 0.1$), but diminishes rapidly in the tank-treading regime. 
The stress contribution from the interaction force [Fig.~\ref{fig_stress_contrib}c, eq.~\eqref{eq_partStresslet_int}] remains moderate in all cases studied, growing not larger than around 20\% of the total stress even in the densest suspension. 
The contributions due to area and volume incompressibility are negligible and we do not show them here.

Interestingly, the relative stress due to shear forces increases with shear rate and decreases with volume fraction, whereas relative bending and interaction stresses show the opposite behavior. 
In contrast to the total stress, the bending stress is sensitive to the \TBTT\ and shows a scaling collapse in the tank-treading regime.

In principle, both short-range repulsive pair-interactions and particle deformability can give rise to non-Newtonian behavior in a suspension \cite{stickel_fluid_2005, guazzelli_physical_2012}.
As Fig.~\ref{fig_stress_contrib}d (solid lines) shows, excluding the interaction forces from the computation of the particle stress has a negligible effect on the effective viscosity.
Rather, we note that the overall behavior of the viscosity is entirely reflecting the contribution due to the elastic shear strain (dashed lines).
Thus, we may conclude that particle shear elasticity is the major source of non-Newtonian behavior in the present model.

\subsection{Yield stress}
\label{sec_yieldstress}

\begin{figure*}[t]\centering
	(a)\includegraphics[width=0.43\linewidth]{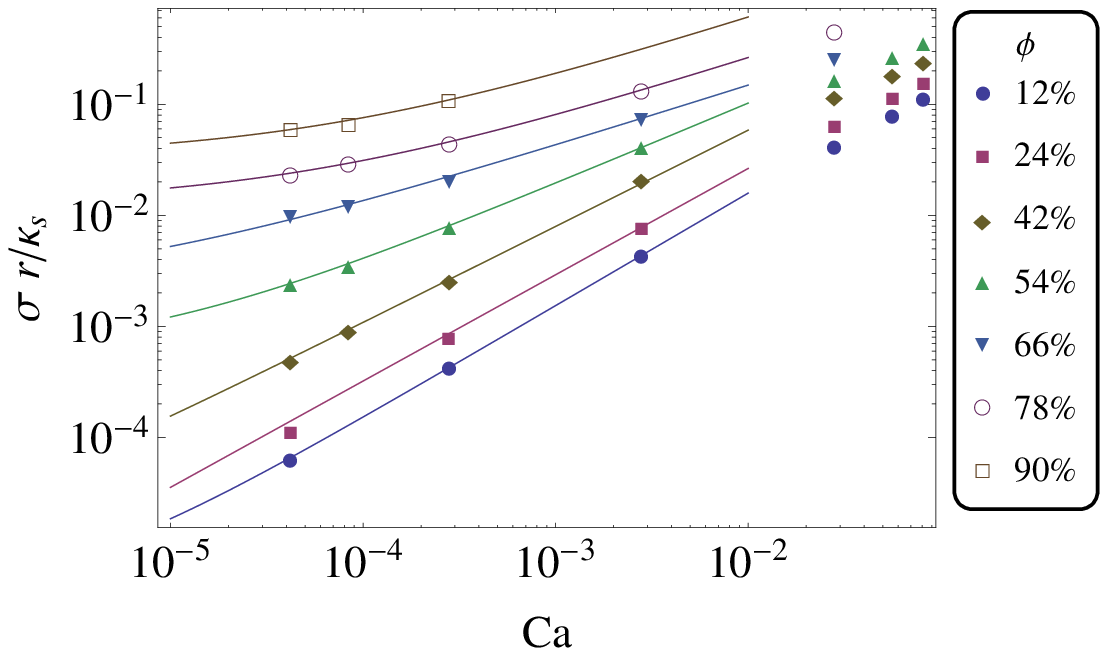}\qquad
	(b)\includegraphics[width=0.41\linewidth]{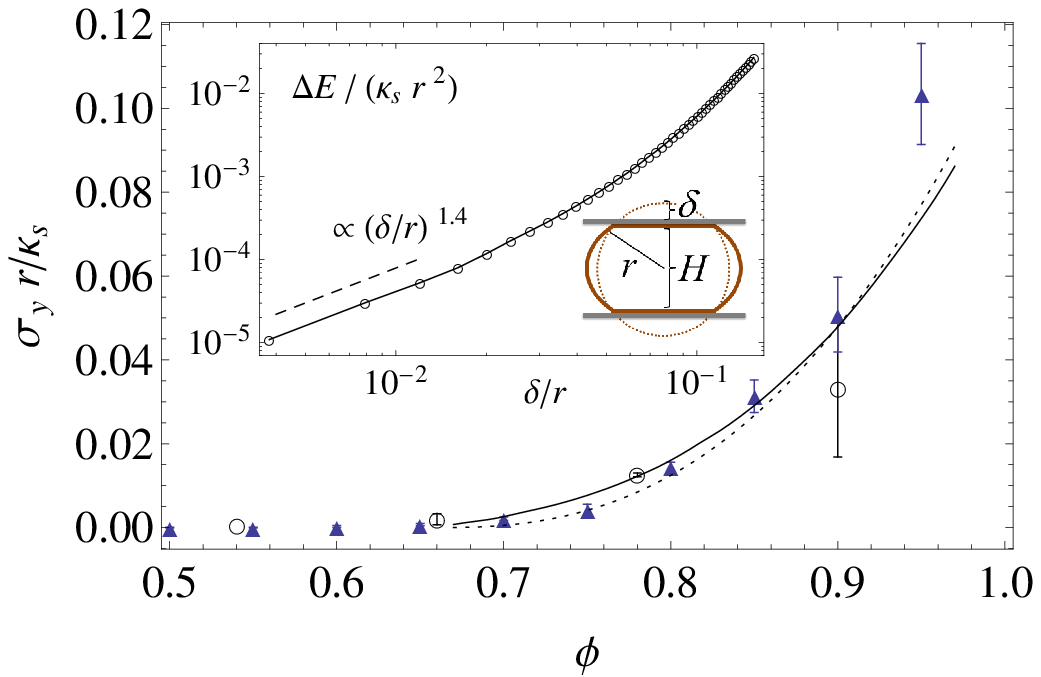}
	\caption{(a) Fits of the Herschel-Bulkley form [eq.~\eqref{eq_HB}] to the suspension shear stress obtained from our simulations. (b) Dependence on volume fraction of the yield stress $\sigma_y$ extracted from experiments \cite{chien_effects_1966} ($\blacktriangle$) and simulations ($\circ$). The data can be empirically described by a power-law (dotted curve) $\sigma_y\propto \Delta\phi^n$, with $n\simeq 2.5$, $\Delta\phi=\phi-\phi_c$ and $\phi_c\simeq 0.66$, see text. The solid curve represents the theoretical model $\sigma_y=c\Delta E/r^3 \Delta\phi^{0.5}$ with the prefactor $c\simeq 4.7$ being the only fit parameter. The inset in (b) shows the increase (over the unstressed state) of the elastic energy, $\Delta E$, of a spherical capsule (radius $r$) compressed by an amount $\delta$ by two walls a distance $H$ apart (see sketch).}
    \label{fig_yieldstress}
\end{figure*}

To capture the rheology in the regime of small shear rates and high volume fractions, we fit a Herschel-Bulkley form \cite{larson_structure_1999}
\beq \hat\sigma = \hat \sigma_y + k\cdot \Ca^p
\label{eq_HB}
\eeq
to the stress as measured by experiments \cite{chien_effects_1966} \footnote{The viscosity values reported in \cite{chien_effects_1966} were given as a function of volume fraction for a number of different shear rates, which we transformed into the representation $\sigma(\dot\gamma)$ via interpolation, cf.\ Fig.~\ref{fig_visc}a.} and our simulations (see Fig.~\ref{fig_yieldstress}a). Here, $\hat\sigma \equiv\sigma r/\mods=\Caeff$ and $\hat \sigma_y\equiv \sigma_y r/\mods$ denote the shear stress and yield stress non-dimensionalized by the scale of the elastic membrane stress ($\mods$ being the shear modulus and $r$ the large semiaxis of the RBC) and $k$ and $p$ are free fit parameters.
The Herschel-Bulkley exponent results as $p= 0.56\pm 0.05$ in simulations and $p= 0.61\pm 0.05$ in experiments. 
The yield stress (Fig.~\ref{fig_yieldstress}b) is found to practically vanish below a critical volume fraction $\phi_c$, which can be associated with the random close packing value $\phi_c\simeq 0.66$ for oblate ellipsoids of the same aspect ratio ($\sim 0.33$) as RBCs \cite{donev_improving_2004}\footnote{In principle, a larger value of $\phi_c$ might be expected due to nematic ordering \cite{kruger_crossover_2013}.}. 

The growth of $\sigma_y$ above $\phi_c$ can be related to the increase of contact energy upon compression of the particles above $\phi_c$ \cite{bolton_rigidity_1990, mason_yielding_1996, coussot_rheophysics_2007, van_hecke_jamming_2010}. 
To gain further insight, we have numerically determined the dependence of the elastic energy $\Delta E$ on the amount of compression (characterized by a indentation parameter $\delta=r-H/2$) for our mechanical model in a simple setup, where a spherical capsule (radius $r$) is compressed by two solid walls a distance $H$ apart (see inset to Fig.~\ref{fig_yieldstress}b). 
We find (inset to Fig.~\ref{fig_yieldstress}b) that the contact energy scales $\propto (\delta/r)^{1.4}$ at small compressions and thus significantly deviates from the naive expectations of Hertzian theory \cite{landau_theory_1986}, which predicts an exponent of 2.5. 
The different scaling can be attributed to the presence of bending elasticity and will be discussed in more details in a future work.
The strong increase of the energy at large compressions is related to the incompressibility of the capsule \cite{siber_many-body_2013}.

In order to transfer these results to a dense suspension, we note that the amount of compression is related to the volume fraction via $\delta/r\simeq 1-(\phi_c/\phi)^{1/3}$ \footnote{Note that $\delta/r\simeq 0.14$ at $\phi=1$, thus the whole range of the measured $\Delta E$ is relevant.}.
The yield stress can be estimated as $\sigma_y=G\gamma_y\simeq \Delta E/\Delta \phi^{0.5}$, where $\Delta\phi\equiv \phi-\phi_c$ and we assumed a scaling of the shear modulus, $G\sim \Delta E/\Delta\phi^{1.5}$, and yield strain, $\gamma_y\sim \Delta\phi$, as typical for disordered solids \cite{mason_yielding_1996, ohern_jamming_2003}.
The resulting function $\sigma_y(\Delta\phi)$ (solid curve in Fig.~\ref{fig_yieldstress}b) fits the simulated and experimental data quite well, with a prefactor of around $4.7$. 
We remark that, empirically, the yield stress can also be described by a simple power law $\sigma_y\propto \Delta\phi^n$, with $n\simeq 2.5\pm 0.5$. While previous experimental studies of the yield stress of RBC suspensions have obtained similar power-law scalings \cite{zydney_constitutive_1991}, the connection to micro-mechanical properties of the cells has not been elucidated.
Note that experiments have often shown a large spread in the extracted values of the yield stress, that is affected partly also by the measurement procedure \cite{picart_human_1998}.

\subsection{Critical jamming scenario}

\begin{figure}[t]\centering
	\includegraphics[width=0.95\linewidth]{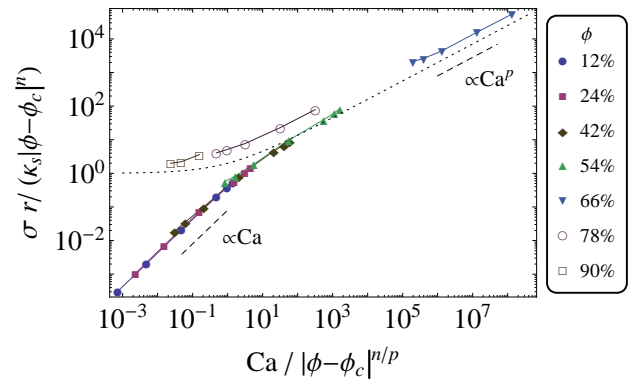}
	\caption{Critical jamming diagram of the rescaled stress vs.\ capillary number obtained from simulations. The dotted curve represents the function $1+x^p$. The yield stress exponent is taken as $n\simeq 2.5$ and the Herschel-Bulkley exponent as $p\simeq 0.55$.}
    \label{fig_critscal}
\end{figure}

The overall rheology of RBC suspensions can be represented in terms of a critical jamming ``phase diagram'', originally introduced in the study of athermal disordered model suspensions \cite{olsson_critical_2007}.
Scaling the viscous stress and the capillary number by (a certain power of) the distance to the jamming point $\phi_c$ results in a remarkable data collapse onto a sub- and supercritical branch, as seen in Fig.~\ref{fig_critscal}.
Note that the critical jamming framework predicts only three distinct rheological regimes, which implies that the strong-deformation regime (corresponding to $\Caeff\gtrsim 0.1$) and yield-stress regime ($\Caeff\lesssim 0.1$ and $\phi\gtrsim \phi_c$) share the same power-law exponents characterizing the shear-thinning of the viscosity, i.e., $q\simeq p-1$, which is consistent with our simulations.
This scenario, however, does not account for the presumed second Newtonian regime at very large capillary numbers observed in experiments \cite{schmid-schonbein_fluid_1969}.

\subsection{Normal stresses}
\begin{figure}[t]\centering
	(a)\includegraphics[width=0.94\linewidth]{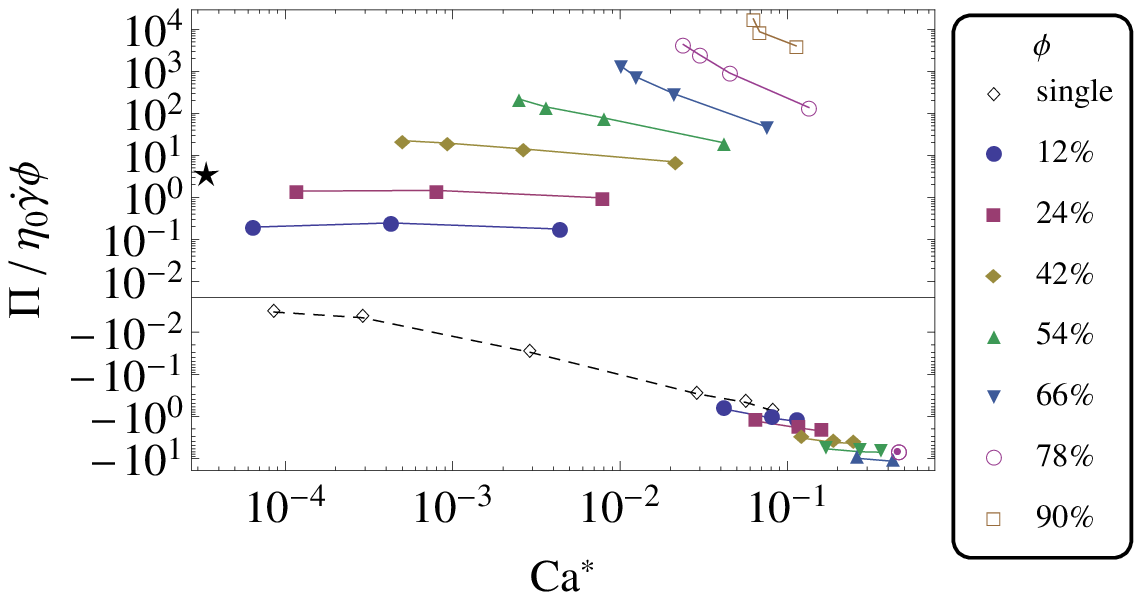}
	(b)\includegraphics[width=0.8\linewidth]{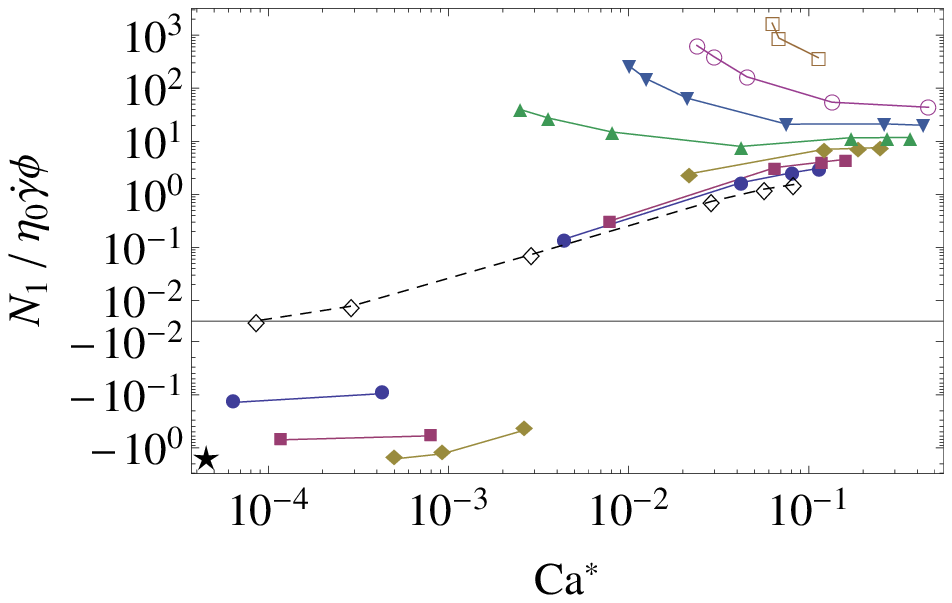}
	(c)\includegraphics[width=0.8\linewidth]{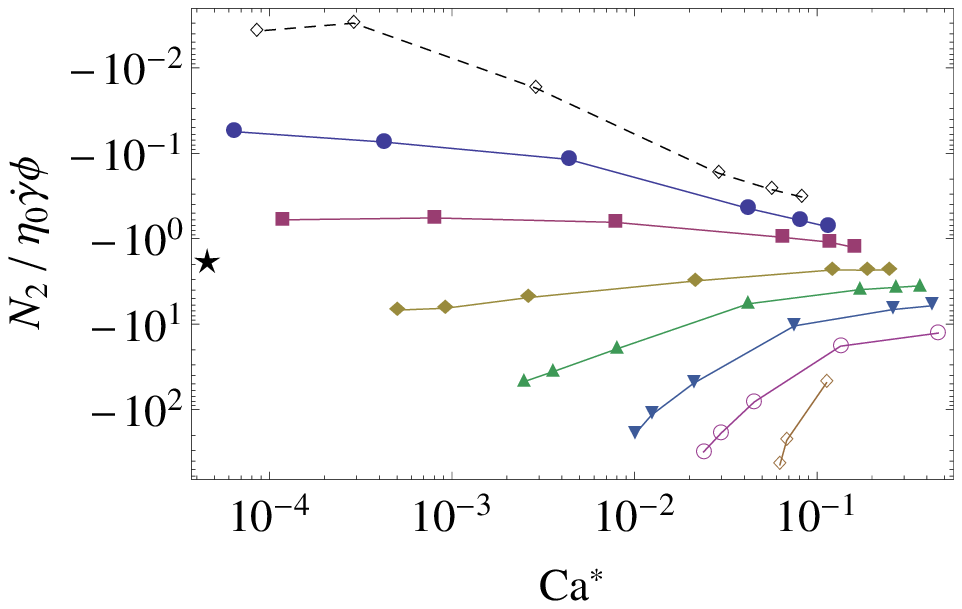}
	\caption{Particle pressure $\Pi$ (a) as well as first and second normal stress differences $N_1, N_2$ (b,c) in RBC suspensions. Note the presence of both positive and negative values of $\Pi$ and $N_1$ in (a,b). The dashed curve corresponds to a single RBC. The asterisk ({$\star$}) marks the value of the corresponding quantity for non-colloidal suspensions of rigid particles at $\phi=40\%$ \cite{sierou_rheology_2002}. The legend in (a) applies to all plots.}
    \label{fig_normstress}
\end{figure}

Fig.~\ref{fig_normstress} shows the results for the particle pressure and normal stress differences [eq.~\eqref{eq_normstress}] obtained from the present simulations. Analogous to the shear stress, reduced quantities are considered by normalizing by the bare shear stress $\eta_0\dot{\gamma}$ and volume fraction $\phi$. 
Consistent with expectations from semi-analytical calculations and numerical simulations of a spherical deformable capsule \cite{ramanujan_deformation_1998, clausen_capsule_2010}, the particle pressure for a single, tumbling RBC (dashed curve in Fig.~\ref{fig_normstress}a) is negative for all capillary numbers.
Hence, the isotropic pressure contribution generated by a single RBC is always tensile in nature.
However, already at $\phi\simeq 12\%$, $\Pi$ becomes positive (compressive) and increases further with volume fraction at low capillary numbers, in agreement with the overall trends reported in \cite{clausen_rheology_2011} for spherical deformable capsules and in \cite{reasor_rheological_2013} for RBC suspensions. A positive particle pressure has also been found for non-colloidal suspensions of rigid spherical particles (the value of $\Pi$ at $\phi=40\%$ is marked by an asterisk in Fig.~\ref{fig_normstress}a), suggesting that a compressive particle pressure in non-dilute suspensions of deformable particles is due to particle interactions.
Interestingly, for effective capillary numbers around $\Ca^*\simeq 0.1$, the reduced particle pressure becomes negative again and, independent of volume fraction, closely matches the behavior of a single RBC. This indicates that, in this regime, single-cell behavior dominates over cell-interactions in $\Pi$. This is not surprising, as, above $\Ca^*\simeq 0.1$, RBCs are in the tank-treading regime \cite{kruger_crossover_2013}, where, in contrast to the tumbling case, encounters with neighboring cells are expected to be much reduced.

The first normal stress difference (Fig.~\ref{fig_normstress}b) shows an interesting bifurcation behavior, in that its small-$\Ca$ limit is negative for volume fractions below $\sim 40\%$ and positive for larger ones.
In both cases, $N_1$ is significantly deviating from its value for a single RBC, signaling again the dominance of cell-cell interactions in the rigid particle limit. However, as far as $N_1$ is concerned, these interactions appear to be of quite different nature for small and large volume fractions. 
Of course, due to the presently limited range of $\Ca^*$, we cannot exclude the possibility that $N_1$ at large volume fractions might become negative at very small $\Ca^*$.
The second normal stress difference (Fig.~\ref{fig_normstress}c) behaves more smoothly and increases in magnitude with increasing volume fraction.
We note that, for small $\Ca^*$ and low volume fractions, the values of $N_1$ and $N_2$ are of comparable magnitude and same sign as in athermal rigid particle suspensions \cite{sierou_rheology_2002}. It is expected that a single RBC in the limit $\Ca\ra 0$ exhibits zero normal stress differences and particle pressure and behaves thus purely Newtonian, as is the case, for instance, for elongated rigid particles \cite{zirnsak_normal_1994}.
A change in the sign of $N_1$ with increasing capillary number has also been observed for spherical deformable capsules \cite{clausen_rheology_2011} at low and intermediate volume fractions ($\phi\lesssim 0.4$), but so far not for RBCs. 
For suspensions of rigid oblate particles, a crossover from negative to positive $N_1$ with increasing volume fraction has been reported in \cite{bertevas_simulation_2010}.

In the limit of large effective capillary numbers (i.e., $\Ca^*\gtrsim 0.1$), both $N_1$ and $N_2$ approach the single particle behavior. Similarly to the particle pressure, this is expected to be a consequence of the \TBTT, which leads to an effective isolation of the suspended cells.
The positive value of $N_1$ in the tank-treading regime reflects the overall tensile structure of normal stresses (i.e., the cell is more intensely stretched along the flow direction than it is compressed in the shear-gradient direction). 
The overall compressive nature of the normal stresses at low capillary numbers is more difficult to interpret, as it is sensitively related to the suspension microstructure.

\section{Effective medium models}
\label{sec_effmed}

\subsection{Conventional approach}
A phenomenological approach to develop some understanding of the behavior of the effective suspension viscosity is provided by effective medium theory \cite{larson_structure_1999}, which has been successfully applied to a variety of different types of suspensions and emulsions, see, e.g., \cite{snabre_rheology_1999, pal_rheology_2003, pal_rheology_2011}. 
The starting point of effective medium theory is the Einstein-type expression for the viscosity of the suspension in the dilute case, as given by
\beq \eta(\phi,\Ca) = \eta_0 + \eta_0 \phi \eta\in(\Ca)\,,
\label{eq_eta_etain_unconf}
\eeq
with $\eta\in(\Ca)$ being the intrinsic viscosity of a single particle (see Fig.~\ref{fig_singlecell_visc}). 
The viscosity of a dense suspension is obtained based on the idea that, upon successively adding particles, the viscosity increment at each step is given by eq.~\eqref{eq_eta_etain_unconf}, with, however, the bare viscosity replaced by the effective viscosity in order to account for the particles already present.
This idea is formalized by rewriting eq.~\eqref{eq_eta_etain_unconf} in differential form,
\beq \frac{\mathrm{d}\eta(\phi,\Ca)}{\mathrm{d}\phi} = \eta(\phi,\Ca) \eta\in\left(\frac{\eta(\phi,\Ca)}{\eta_0} \Ca\right)\,,
\label{eq_effmed_visc_unconf}
\eeq
which is to be solved subject to the initial condition $\eta(\phi=0,\Ca) = \eta_0$. 
The replacement of the bare by the effective capillary number, $\Ca^*=(\eta/\eta_0) \Ca$ in eq.~\eqref{eq_effmed_visc_unconf} is central to the application of the model to a soft-particle suspension and encapsulates the notion that viscosity is dominated by capsule deformation, hence $\Caeff$ (see \cite{kruger_crossover_2013} and Fig.~\ref{fig_singlecell_visc}). 

In order to account for excluded volume effects with increasing $\phi$, refinements of the simple effective medium equation~\eqref{eq_effmed_visc_unconf} have been proposed \cite{krieger_mechanism_1959} in which the volume increment is $d\phi/(1-\phi/\phi_m$) rather than $d\phi$. Here, $\phi_m$ is the maximum packing fraction fraction at which the viscosity is supposed to diverge. In the case of hard particles, this is typically the volume fraction of random close packing. However, when particles are deformable, packing fractions close to 1 are possible (see Fig.~\ref{fig_visc}), which suggests to take $\phi_m=1$ in the present case.
Taken together, a refined effective medium equation can be proposed as \cite{pal_rheology_2003}
\beq \frac{\mathrm{d}\eta(\phi,\Ca)}{\mathrm{d}\phi} = \frac{\eta(\phi,\Ca) }{1-\phi}\eta\in\left(\frac{\eta(\phi,\Ca)}{\eta_0} \Ca\right)\,.
\label{eq_effmed2_visc_unconf}
\eeq
It is useful to note that, in the limit $\Ca\ra 0$, the above effective medium models predict viscosities that are independent of the constitutive law represented by $\eta\in(\Ca)$: from eq.~\eqref{eq_effmed_visc_unconf}, an exponential dependence on volume fraction results,
\beq \eta(\phi,\Ca=0) = \eta_0 \exp(\eta\st{in,0} \phi)\,,
\label{eq_lim_visceff1}
\eeq
with $\eta\st{in,0}\equiv \eta\in(\Ca=0)$, whereas eq.~\eqref{eq_effmed2_visc_unconf} predicts a power-law divergence at $\phi_m=1$,
\beq \eta(\phi,\Ca=0) = \eta_0 \left(1-\phi\right)^{-\eta\st{in,0}} \,,
\label{eq_lim_visceff2}
\eeq
which corresponds in fact to the well-known viscosity formula of Krieger and Dougherty \cite{krieger_mechanism_1959}.

\begin{figure*}[t]\centering
	(a)\includegraphics[width=0.45\linewidth]{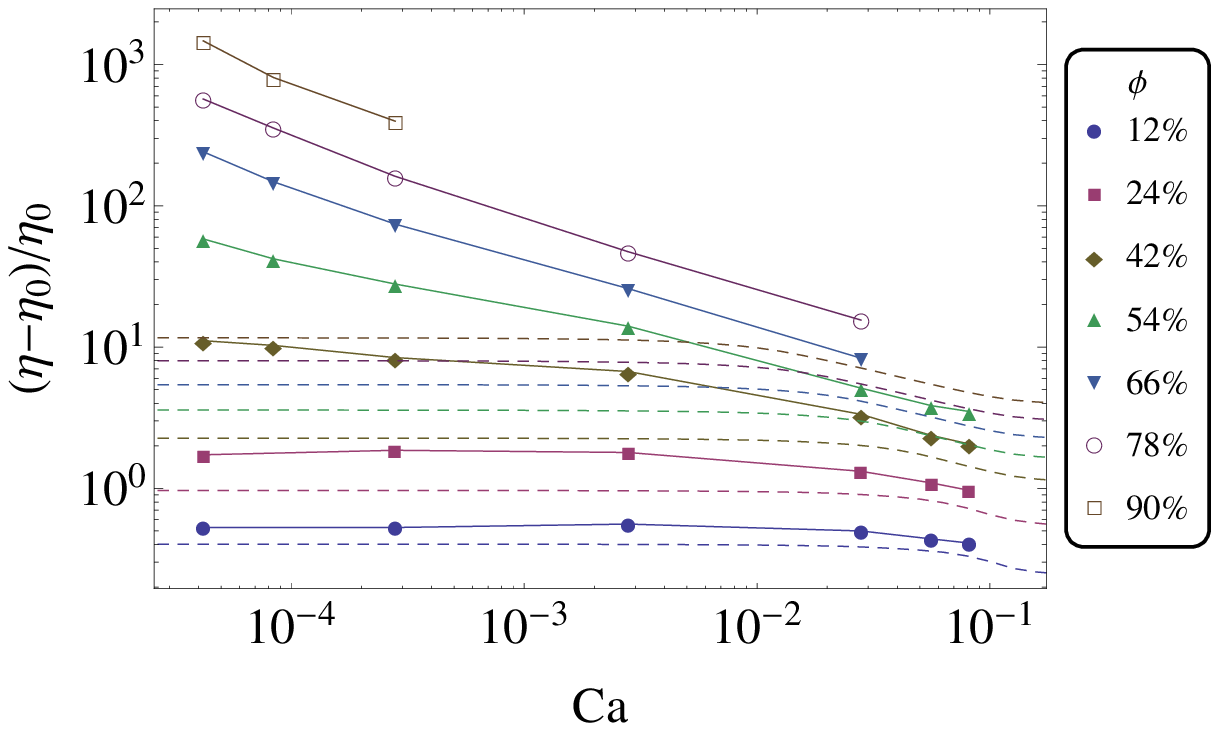}\qquad
	(b)\includegraphics[width=0.45\linewidth]{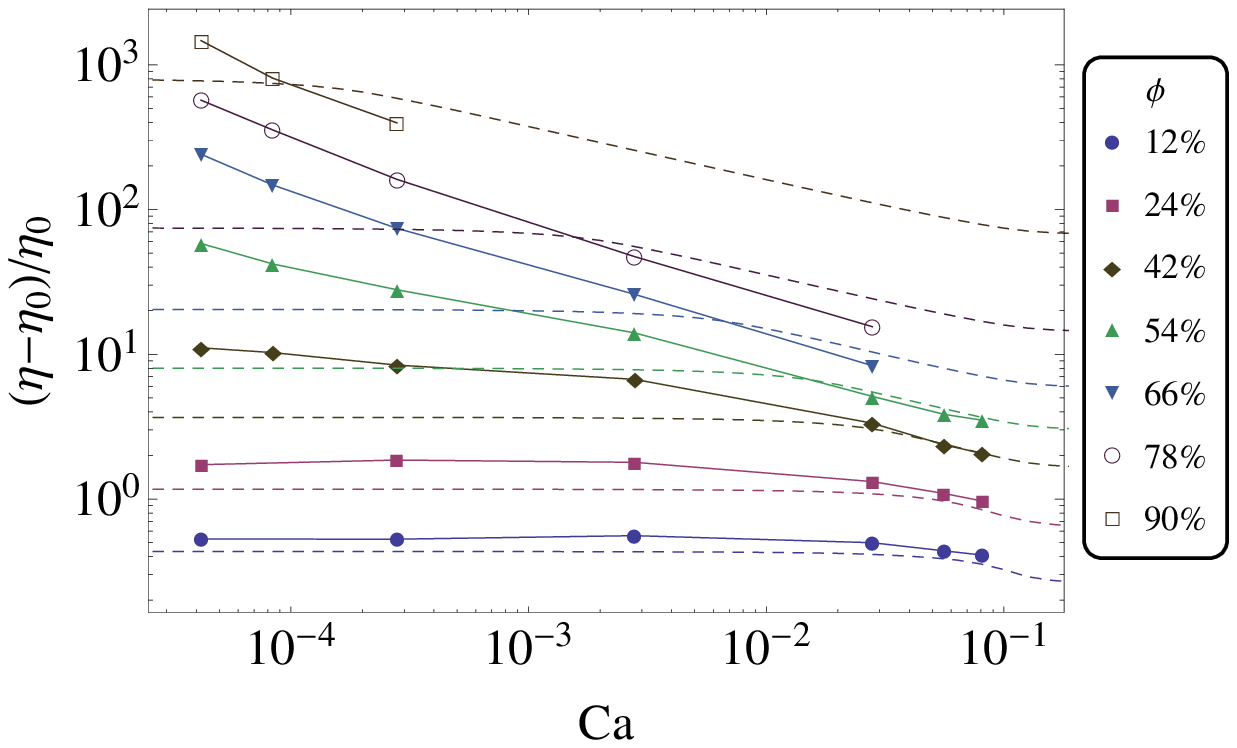}
	\caption{Predictions (dashed curves) of the effective medium models of (a) eq.~\eqref{eq_effmed_visc_unconf} and (b) eq.~\eqref{eq_effmed2_visc_unconf}. In both cases, analytical predictions deviate strongly from simulation results (connected symbols) and also fail to predict the characteristic shear-thinning exponent.}
    \label{fig_effmed_unconf}
\end{figure*}

To evaluate the effective medium equations, we shall use, for definiteness, in the following the intrinsic viscosity of an RBC that has its major plane perpendicular to the shearing plane (in which case we find $\eta\in(\Ca=0)\simeq 2.8$, see Fig.~\ref{fig_singlecell_visc}a). 
This corresponds to the most probable alignment of cells in suspension \cite{gross_inprep}. 
The viscosity curve is furthermore smoothly extrapolated towards smaller and larger capillary numbers by assuming a constant $\eta\in$ \footnote{The effective medium equations are found to be not very sensitive to the numerical details of the continuation.}.
Fig.~\ref{fig_effmed_unconf}a,b show the results of the integration of eqs.~\eqref{eq_effmed_visc_unconf} and ~\eqref{eq_effmed2_visc_unconf}, respectively.
Although deviations between simulation data and theory appear already at small volume fractions, we note that the effective medium model captures some of the aspects of the suspension rheology -- in particular, the shift of the shear-thinning regime towards smaller $\Ca$ with increasing volume fraction and the approximately exponential dependence of the viscosity on volume fraction. 
We note that the quality of these results is comparable to previous effective medium approaches to model the viscosity of blood based on a constitutive relation of a single isolated cell \cite{pal_rheology_2003}. 

\subsection{Confinement effects}
In a dense suspension, hydrodynamic interactions between particles generate additional stresses which significantly alter the effective viscosity over its value in the dilute case \cite{batchelor_determination_1972}. 
If the particles are deformable, the situation is complicated by the fact that surface stresses and particle deformation are coupled.
This aspect is most eminent at intermediate capillary numbers, where shear-thinning and hence nonlinear deformation effects are strong. 
This regime is also the physiologically most relevant one for blood flow.
Conventional effective medium models account for particle interactions in a phenomenological way by assuming that the suspension becomes overall more viscous if particles are added.
As the results of the previous section indicate, this approach obviously fails to capture all relevant effects in the present case.
It appears that the mutual confinement of the particles in a dense suspension influences the single-particle viscosity in a way not accounted for by the simple models of eqs.~\eqref{eq_effmed_visc_unconf} or \eqref{eq_effmed2_visc_unconf}.

\begin{figure}[t]\centering
	\includegraphics[width=0.82\linewidth]{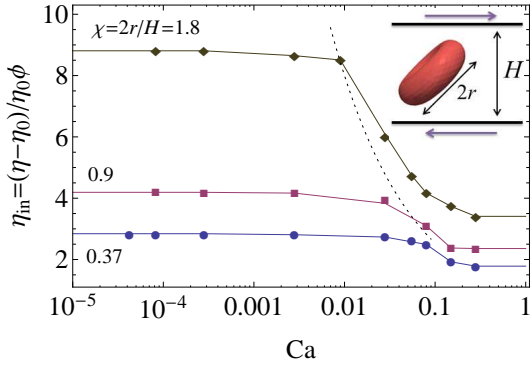}\quad
	\caption{Intrinsic viscosity $\eta\in$ of a single RBC (large radius $r$) in shear flow confined by solid walls (separation $H$) for various levels of confinement $\chi=2r/H$. 
	Simulated data (symbols) are extrapolated toward small and large $\Ca$ by assuming a constant $\eta\in$. For comparison, the power-law $\eta\in\propto \Ca^{-0.5}$ (dotted curve) that characterizes the shear-thinning behavior of a dense suspension (cf.~Fig.~\ref{fig_visc}a) is included.}
    \label{fig_visc_sc_conf}
\end{figure}

In order to understand the principal effects on the visosity induced by confinement of a deformable particle, we first study the situation where a single RBC (large radius $r$) in shear flow is confined by solid walls (a distance $H$ apart), see inset to Fig.~\ref{fig_visc_sc_conf}.
The resulting intrinsic viscosity $\eta\in$ for different levels of confinement $\chi \equiv 2r/H$ is shown in the main panel of Fig.~\ref{fig_visc_sc_conf}. 
As before, the major plane of the cell is oriented perpendicular to the shearing plane.
Interestingly, a stronger confinement not only leads to an increased viscosity, but also enhances the shear-thinning regime. 
In the case of rigid particle suspensions, the viscosity increase under confinement is well known \cite{guth_untersuchungen_1936, davit_intriguing_2008, swan_particle_2010, peyla_new_2011, sangani_roles_2011}.
Note that the shear-thinning behavior for a single cell is very different from the power-law $\eta \propto \Ca^{-0.5}$ (dashed curve in Fig.~\ref{fig_visc_sc_conf}) that applies to a dense suspension (cf.\ Fig.~\ref{fig_visc}a).
We have not included viscosity values for confinements $\chi$ close to 1, since in that case the cell assumes a static orientation almost perpendicular to the flow, leading to a spurious rise of the effective viscosity. 
The problem does not occur for very strong confinements (here, $\chi=1.8$), however. 
We finally remark that effects of confining walls on the dynamics of vesicles and RBCs have been studied in \cite{kaoui_how_2012}.

\begin{figure}[t]\centering
	\includegraphics[width=0.95\linewidth]{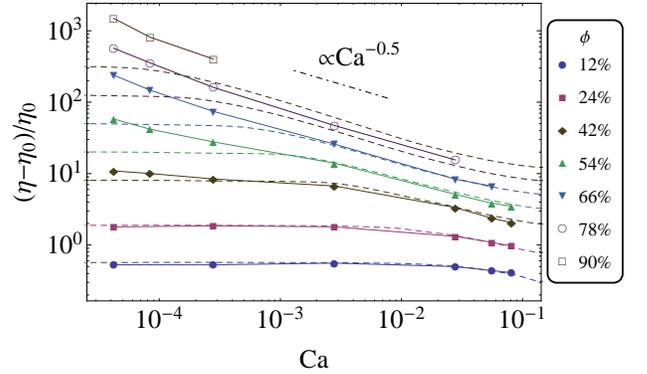}
	\caption{Predictions of the effective medium model including confinement effects, eq.~\eqref{eq_effmed_visc} (dashed curves), in comparison to the simulation results (connected symbols).}
    \label{fig_effmed_conf}
\end{figure}

Returning to the case of a bulk suspension, we may resort -- in lack of a detailed theoretical model -- to a somewhat crude approach and assume that the essential effects of confinement due to neighboring particles are similar to the situation of wall-induced confinement.
Note that we focus here exclusively on confinement effects on the viscosity in the bulk, ignoring phenomena such as the Fahraeus-Lindqvist effect \cite{faahraeus_viscosity_1931}, which would require an extension of the present model.
To proceed, we propose a modified effective medium equation 
\beq \frac{\mathrm{d}\eta(\phi,\Ca)}{\mathrm{d}\phi} = \eta(\phi,\Ca) \eta\in\left(\frac{\eta(\phi,\Ca)}{\eta_0} \Ca, \chi(\phi)\right)\,,
\label{eq_effmed_visc}
\eeq
where $\eta\in$ is now the intrinsic single-cell viscosity displayed in Fig.~\ref{fig_visc_sc_conf}, linearly interpolated over the missing intermediate confinement values. 

If $N$ particles are homogeneously suspended in a volume $V$, the typical volume available to each particle can be estimated as $d^3 = V/N = v_c/\phi$, with $v_c\simeq 1.6 r^3$ being the volume of an RBC.
To obtain a mapping relation between confinement and volume fraction, we take $H\simeq d$, which gives
\beq \chi=2 c r(\phi/v_c)^{1/3}\,.
\label{eq_conf_mapping}
\eeq 
Here, $c\simeq 1.15$ is a free parameter that has been adjusted to obtain the best agreement with our simulation results. 
Note that $\chi\simeq 1$ already at $\phi\simeq 0.15$, thus confinement effects are relevant for all volume fractions studied here (cf.~\cite{levant_characteristic_2012}).

Solving eq.~\eqref{eq_effmed_visc} subject to the initial condition $\eta(\phi=0,\Ca) = \eta_0$ results in the effective suspension viscosity represented by the dashed curves in Fig.~\ref{fig_effmed_conf}.
Remarkably, despite the rather simplistic approach to incorporate confinement effects, we obtain an impressive agreement over a wide range of volume fractions and capillary numbers. 
In particular, the effective power-law $\eta\propto \Ca^{-0.5}$, characterizing the shear-thinning of the viscosity over a large range of capillary numbers (Fig.~\ref{fig_visc}a), is correctly predicted and is seen here to emerge in a nontrivial way from the shear-thinning behavior of a single cell, which obeys a different Ca-dependence (see Fig.~\ref{fig_visc_sc_conf}).
Consistent with the critical jamming scenario, the value of the above exponent matches the expectation from the Herschel-Bulkley fits, $p-1$ with $p\simeq 0.56\pm 0.05$.
Note that no fitting parameters are involved here, except for the choice of the mapping relation between $\phi$ and $\chi$ [eq.~\eqref{eq_conf_mapping}].

At small capillary numbers, the suspension viscosity obtained from the effective medium models necessarily exhibits a Newtonian plateau, as this characteristic is already present in the intrinsic viscosity of a single particle.
The present models can thus not capture the apparent divergence of the suspension viscosity in the limit $\Ca\ra 0$ at large $\phi$.

In order to gain further insight into the principal behavior of the effective medium eq.~\eqref{eq_effmed_visc}, consider a simple toy model where the single-cell viscosity follows a pure power-law behavior of the form $\eta\in(\Ca) = k \Ca^{-x} \phi^n$, with $k$ being a constant and $x$ and $n$ some exponents. 
For instance, $n=1/3$ according to eq.~\eqref{eq_conf_mapping}, while the shear-thinning of the viscosity for $\chi=1.8$ in Fig.~\ref{fig_visc_sc_conf} can be described by $x\simeq 0.2-0.3$.
Now, eq.~\eqref{eq_effmed_visc} can be easily integrated to give 
\beq \eta(\phi,\Ca) = \eta_0 \left(1 + \frac{k x \phi^{1+n}}{1+n} \Ca^{-x} \right)^{1/x}\,.
\label{eq_lim2_visceff1}
\eeq
Note that, in the limit $x\ra 0$, $n=0$, eq.~\eqref{eq_lim2_visceff1} reduces to the $\Ca$-independent expression of eq.~\eqref{eq_lim_visceff1} 
with $\eta\st{in,0}=k$.
More interesting is the behavior at finite $x$, where eq.~\eqref{eq_lim2_visceff1} approaches a constant for large Ca and a unique power-law $\eta \sim \Ca^{-1}$ in the limit $\Ca\ra 0$. 
From eq.~\eqref{eq_lim2_visceff1} we may also note that, since the crossover to the limiting behavior at $\Ca\ra 0$ is determined by the prefactor of the term $\Ca^{-x}$, the steepness of the effective viscosity curves increases with $\phi$ but decreases with $n$ (at fixed Ca).
Obviously, any finite exponent $x$ characterizing the single-cell viscosity will be renormalized by the effective medium equation and give rise to a range of effective shear-thinning exponents of the suspension viscosity $\eta$. 
Thus, the emergence of the rather robust power-law $\eta\propto \Ca^{-0.5}$ in the present case is a consequence of the specific Ca- and $\chi$-dependencies of the single-cell viscosity.

\section{Summary and discussion}
In the present work, we have studied a suspension of aggregation-free red blood cells, focusing on the connection between micro-mechanical properties of the capsules and the macroscopic rheology.
The capsules are modeled as incompressible elastic membranes with a certain shear and bending stiffness.
Thermal fluctuations are absent and particles interact only via hydrodynamics and short-range repulsive forces. The latter are essentially included to improve numerical stability and their influence on the overall rheology is found to be weak. 
Remarkably, the complex shape and rotational dynamics of an RBC does not show up prominently in the macro-scale rheology, which, rather, is determined by the shear elasticity of the particles and the distance to the jamming point. 

The shear viscosity is in good agreement with previous experimental studies \cite{chien_effects_1966} and exhibits three distinct regimes: at small capillary numbers, a Newtonian plateau is present at low volume fractions, which goes over into a yield stress regime at high volume fractions; for large capillary numbers, the viscosity is strongly shear-thinning, following a power-law $\eta\propto \Ca^{q}$, with $q\simeq -0.5$.
Consistent with this behavior and as expected from the critical jamming scenario, the Herschel-Bulkley fits in the yield stress regime are described by an exponent $p\simeq 1+q$.
We remark that a Herschel-Bulkley exponent of 1/2 has also been observed in simulations of model foams \cite{langlois_rheological_2008, chaudhuri_inhomogeneous_2012, mansard_molecular_2013} as well as in generic elasto-plastic models \cite{bocquet_kinetic_2009}, and has been predicted theoretically for soft particle pastes based on a coupling between lubrication flow and elasticity \cite{denkov_viscous_2008, seth_micromechanical_2011}. 
The yield stress is found to be consistent with a model based on elastic contact interactions between cells, assuming a typical scaling of shear modulus and yield strain in amorphous solids.
We found that simple effective medium models based on the intrinsic viscosity of a free particle did not provide a quantitatively satisfactory description of the suspension viscosity.
This is interpreted as to point to the importance of confinement effects, which, for the case of a capsule in wall-bounded shear flow, were shown to not only increase the overall magnitude of the viscosity, but also to enhance shear-thinning. 
Based on a modified effective medium model, the power-law exponent $q$ characterizing the shear-thinning of the suspension viscosity can then be understood as a consequence of a ``renormalization'' due to collective effects of the shear-thinning behavior of single confined cell. 

In contrast to the shear stress, the normal stresses appear to be a much more sensitive rheological probe of the competition between collective interactions and single-cell properties.
At small capillary numbers, the particle pressure and normal stress differences behave in a similar way to athermal rigid particle suspensions, while, at larger capillary numbers, they follow the behavior of a single cell.
An -- so far unexplained -- exception occurs for the first normal stress difference, which, at low effective capillary numbers, crosses over from negative to positive values with increasing volume fraction. 
The particle pressure displays a dramatic sign change around the critical effective capillary number characterizing the \TBTT\ and scales independent of volume fraction in the tank-treading regime. 
Shear-rate dependent particle pressure and normal stress differences can induce particle migration and lead to a coupling between concentration and flow field \cite{schmitt_shear-induced_1995}. The consequences of such behavior may include shear-banding and ``rod-climbing'' effects \cite{malkin_rheology:_2006} and have recently received strong interested in the field of soft glassy rheology \cite{besseling_shear_2010, mandal_heterogeneous_2012}. In the context of suspensions of deformable particles, these issues are largely unexplored.

Our work is hoped to contribute to a better understanding of the rheology of blood and other soft-particle suspensions beyond purely phenomenological relations \cite{yilmaz_critical_2008, vlahovska_vesicles_2009}. 

\acknowledgements{
We thank C.~Heussinger for valuable discussions and S.\ K.\ Lanka for his contributions to the code development. This work is financially supported by the DFG-project Va205/5-2. We are also grateful for the computational time granted by the J\"ulich Supercomputing Centre (Project ESMI17). 
}


\begin{thebibliography}{107}
\expandafter\ifx\csname natexlab\endcsname\relax\def\natexlab#1{#1}\fi
\expandafter\ifx\csname bibnamefont\endcsname\relax
  \def\bibnamefont#1{#1}\fi
\expandafter\ifx\csname bibfnamefont\endcsname\relax
  \def\bibfnamefont#1{#1}\fi
\expandafter\ifx\csname citenamefont\endcsname\relax
  \def\citenamefont#1{#1}\fi
\expandafter\ifx\csname url\endcsname\relax
  \def\url#1{\texttt{#1}}\fi
\expandafter\ifx\csname urlprefix\endcsname\relax\def\urlprefix{URL }\fi
\providecommand{\bibinfo}[2]{#2}
\providecommand{\eprint}[2][]{\url{#2}}

\bibitem[{\citenamefont{Stickel and Powell}(2005)}]{stickel_fluid_2005}
\bibinfo{author}{\bibfnamefont{J.~J.} \bibnamefont{Stickel}} \bibnamefont{and}
  \bibinfo{author}{\bibfnamefont{R.~L.} \bibnamefont{Powell}},
  \bibinfo{journal}{Annu. Rev. Fluid Mech.} \textbf{\bibinfo{volume}{37}},
  \bibinfo{pages}{129} (\bibinfo{year}{2005}).

\bibitem[{\citenamefont{Coussot}(2007)}]{coussot_rheophysics_2007}
\bibinfo{author}{\bibfnamefont{P.}~\bibnamefont{Coussot}},
  \bibinfo{journal}{Soft Matter} \textbf{\bibinfo{volume}{3}},
  \bibinfo{pages}{528} (\bibinfo{year}{2007}).

\bibitem[{\citenamefont{Schall and van Hecke}(2010)}]{schall_shear_2010}
\bibinfo{author}{\bibfnamefont{P.}~\bibnamefont{Schall}} \bibnamefont{and}
  \bibinfo{author}{\bibfnamefont{M.}~\bibnamefont{van Hecke}},
  \bibinfo{journal}{Annu. Rev. Fluid Mech.} \textbf{\bibinfo{volume}{42}},
  \bibinfo{pages}{67} (\bibinfo{year}{2010}).

\bibitem[{\citenamefont{Larson}(1999)}]{larson_structure_1999}
\bibinfo{author}{\bibfnamefont{R.~G.} \bibnamefont{Larson}},
  \emph{\bibinfo{title}{The Structure and Rheology of Complex Fluids}}
  (\bibinfo{publisher}{Oxford University Press}, \bibinfo{year}{1999}).

\bibitem[{\citenamefont{Pal}(2007)}]{pal_rheology_2007}
\bibinfo{author}{\bibfnamefont{R.}~\bibnamefont{Pal}},
  \emph{\bibinfo{title}{Rheology of particulate dispersions and composites}}
  (\bibinfo{publisher}{{CRC} Press}, \bibinfo{address}{Boca Raton, {FL}},
  \bibinfo{year}{2007}).

\bibitem[{\citenamefont{Popel and Johnson}(2005)}]{popel_microcirculation_2005}
\bibinfo{author}{\bibfnamefont{A.~S.} \bibnamefont{Popel}} \bibnamefont{and}
  \bibinfo{author}{\bibfnamefont{P.~C.} \bibnamefont{Johnson}},
  \bibinfo{journal}{Ann. Rev. Fluid Mech.} \textbf{\bibinfo{volume}{37}},
  \bibinfo{pages}{43} (\bibinfo{year}{2005}).

\bibitem[{\citenamefont{Yilmaz and Gundogdu}(2008)}]{yilmaz_critical_2008}
\bibinfo{author}{\bibfnamefont{F.}~\bibnamefont{Yilmaz}} \bibnamefont{and}
  \bibinfo{author}{\bibfnamefont{M.~Y.} \bibnamefont{Gundogdu}},
  \bibinfo{journal}{Korea-Aust. Rheol. J.} \textbf{\bibinfo{volume}{20}},
  \bibinfo{pages}{197} (\bibinfo{year}{2008}).

\bibitem[{\citenamefont{Pozrikidis}(2010)}]{pozrikidis_computational_2010}
\bibinfo{author}{\bibfnamefont{C.}~\bibnamefont{Pozrikidis}},
  \emph{\bibinfo{title}{Computational Hydrodynamics of Capsules and Biological
  Cells}} (\bibinfo{publisher}{{CRC} Press}, \bibinfo{year}{2010}).

\bibitem[{\citenamefont{Barth\`{e}s-Biesel}(2011)}]{barthes-biesel_modeling_2011}
\bibinfo{author}{\bibfnamefont{D.}~\bibnamefont{Barth\`{e}s-Biesel}},
  \bibinfo{journal}{Curr. Opin. Coll. Int. Sci.} \textbf{\bibinfo{volume}{16}},
  \bibinfo{pages}{3} (\bibinfo{year}{2011}).

\bibitem[{\citenamefont{Misbah}(2012)}]{misbah_vesicles_2012}
\bibinfo{author}{\bibfnamefont{C.}~\bibnamefont{Misbah}}, \bibinfo{journal}{J.
  Phys.: Conf. Ser.} \textbf{\bibinfo{volume}{392}}, \bibinfo{pages}{012005}
  (\bibinfo{year}{2012}).

\bibitem[{\citenamefont{Yaron and Gal-Or}(1972)}]{yaron_viscous_1972}
\bibinfo{author}{\bibfnamefont{I.}~\bibnamefont{Yaron}} \bibnamefont{and}
  \bibinfo{author}{\bibfnamefont{B.}~\bibnamefont{Gal-Or}},
  \bibinfo{journal}{Rheologica Acta} \textbf{\bibinfo{volume}{11}},
  \bibinfo{pages}{241} (\bibinfo{year}{1972}).

\bibitem[{\citenamefont{Choi and Schowalter}(1975)}]{choi_rheological_1975}
\bibinfo{author}{\bibfnamefont{S.~J.} \bibnamefont{Choi}} \bibnamefont{and}
  \bibinfo{author}{\bibfnamefont{W.~R.} \bibnamefont{Schowalter}},
  \bibinfo{journal}{Phys. Fluids} \textbf{\bibinfo{volume}{18}},
  \bibinfo{pages}{420} (\bibinfo{year}{1975}).

\bibitem[{\citenamefont{Brennen}(1975)}]{brennen_concentrated_1975}
\bibinfo{author}{\bibfnamefont{C.}~\bibnamefont{Brennen}},
  \bibinfo{journal}{Canad. J. Chem. Eng.} \textbf{\bibinfo{volume}{53}},
  \bibinfo{pages}{126} (\bibinfo{year}{1975}).

\bibitem[{\citenamefont{Pal}(2011)}]{pal_rheology_2011}
\bibinfo{author}{\bibfnamefont{R.}~\bibnamefont{Pal}}, \bibinfo{journal}{Curr.
  Opin. Coll. Int.} \textbf{\bibinfo{volume}{16}}, \bibinfo{pages}{41}
  (\bibinfo{year}{2011}).

\bibitem[{\citenamefont{Eggleton and Popel}(1998)}]{eggleton_large_1998}
\bibinfo{author}{\bibfnamefont{C.~D.} \bibnamefont{Eggleton}} \bibnamefont{and}
  \bibinfo{author}{\bibfnamefont{A.~S.} \bibnamefont{Popel}},
  \bibinfo{journal}{Phys. Fluids} \textbf{\bibinfo{volume}{10}},
  \bibinfo{pages}{1834} (\bibinfo{year}{1998}).

\bibitem[{\citenamefont{Zhang et~al.}(2007)\citenamefont{Zhang, Johnson, and
  Popel}}]{zhang_immersed_2007}
\bibinfo{author}{\bibfnamefont{J.}~\bibnamefont{Zhang}},
  \bibinfo{author}{\bibfnamefont{P.~C.} \bibnamefont{Johnson}},
  \bibnamefont{and} \bibinfo{author}{\bibfnamefont{A.~S.} \bibnamefont{Popel}},
  \bibinfo{journal}{Phys. Biol.} \textbf{\bibinfo{volume}{4}},
  \bibinfo{pages}{285} (\bibinfo{year}{2007}).

\bibitem[{\citenamefont{Dupin et~al.}(2007)\citenamefont{Dupin, Halliday, Care,
  Alboul, and Munn}}]{dupin_modeling_2007}
\bibinfo{author}{\bibfnamefont{M.~M.} \bibnamefont{Dupin}},
  \bibinfo{author}{\bibfnamefont{I.}~\bibnamefont{Halliday}},
  \bibinfo{author}{\bibfnamefont{C.~M.} \bibnamefont{Care}},
  \bibinfo{author}{\bibfnamefont{L.}~\bibnamefont{Alboul}}, \bibnamefont{and}
  \bibinfo{author}{\bibfnamefont{L.~L.} \bibnamefont{Munn}},
  \bibinfo{journal}{Phys. Rev. E} \textbf{\bibinfo{volume}{75}},
  \bibinfo{pages}{066707} (\bibinfo{year}{2007}).

\bibitem[{\citenamefont{Bagchi}(2007)}]{bagchi_mesoscale_2007}
\bibinfo{author}{\bibfnamefont{P.}~\bibnamefont{Bagchi}},
  \bibinfo{journal}{Biophys. J.} \textbf{\bibinfo{volume}{92}},
  \bibinfo{pages}{1858} (\bibinfo{year}{2007}).

\bibitem[{\citenamefont{Doddi and Bagchi}(2009)}]{doddi_three-dimensional_2009}
\bibinfo{author}{\bibfnamefont{S.~K.} \bibnamefont{Doddi}} \bibnamefont{and}
  \bibinfo{author}{\bibfnamefont{P.}~\bibnamefont{Bagchi}},
  \bibinfo{journal}{Phys. Rev. E} \textbf{\bibinfo{volume}{79}},
  \bibinfo{pages}{046318} (\bibinfo{year}{2009}).

\bibitem[{\citenamefont{{MacMeccan} et~al.}(2009)\citenamefont{{MacMeccan},
  Clausen, Neitzel, and Aidun}}]{macmeccan_simulating_2009}
\bibinfo{author}{\bibfnamefont{R.~M.} \bibnamefont{{MacMeccan}}},
  \bibinfo{author}{\bibfnamefont{J.~R.} \bibnamefont{Clausen}},
  \bibinfo{author}{\bibfnamefont{G.~P.} \bibnamefont{Neitzel}},
  \bibnamefont{and} \bibinfo{author}{\bibfnamefont{C.~K.} \bibnamefont{Aidun}},
  \bibinfo{journal}{J. Fluid Mech.} \textbf{\bibinfo{volume}{618}},
  \bibinfo{pages}{13} (\bibinfo{year}{2009}).

\bibitem[{\citenamefont{Clausen and Aidun}(2010)}]{clausen_capsule_2010}
\bibinfo{author}{\bibfnamefont{J.~R.} \bibnamefont{Clausen}} \bibnamefont{and}
  \bibinfo{author}{\bibfnamefont{C.~K.} \bibnamefont{Aidun}},
  \bibinfo{journal}{Phys. Fluids} \textbf{\bibinfo{volume}{22}},
  \bibinfo{pages}{123302} (\bibinfo{year}{2010}).

\bibitem[{\citenamefont{Clausen et~al.}(2011)\citenamefont{Clausen, Reason, and
  Aidun}}]{clausen_rheology_2011}
\bibinfo{author}{\bibfnamefont{J.~R.} \bibnamefont{Clausen}},
  \bibinfo{author}{\bibfnamefont{D.~A.} \bibnamefont{Reason}},
  \bibnamefont{and} \bibinfo{author}{\bibfnamefont{C.~K.} \bibnamefont{Aidun}},
  \bibinfo{journal}{J. Fluid Mech.} \textbf{\bibinfo{volume}{685}},
  \bibinfo{pages}{1} (\bibinfo{year}{2011}).

\bibitem[{\citenamefont{Reasor et~al.}(2013)\citenamefont{Reasor, Clausen, and
  Aidun}}]{reasor_rheological_2013}
\bibinfo{author}{\bibfnamefont{D.~A.~J.} \bibnamefont{Reasor}},
  \bibinfo{author}{\bibfnamefont{J.~R.} \bibnamefont{Clausen}},
  \bibnamefont{and} \bibinfo{author}{\bibfnamefont{C.~K.} \bibnamefont{Aidun}},
  \bibinfo{journal}{J. Fluid Mech.} \textbf{\bibinfo{volume}{726}},
  \bibinfo{pages}{497} (\bibinfo{year}{2013}).

\bibitem[{\citenamefont{Schmid-Sch\"onbein and
  Wells}(1969)}]{schmid-schonbein_fluid_1969}
\bibinfo{author}{\bibfnamefont{H.}~\bibnamefont{Schmid-Sch\"onbein}}
  \bibnamefont{and} \bibinfo{author}{\bibfnamefont{R.}~\bibnamefont{Wells}},
  \bibinfo{journal}{Science} \textbf{\bibinfo{volume}{165}},
  \bibinfo{pages}{288} (\bibinfo{year}{1969}).

\bibitem[{\citenamefont{Abkarian et~al.}(2007)\citenamefont{Abkarian, Faivre,
  and Viallat}}]{abkarian_swinging_2007}
\bibinfo{author}{\bibfnamefont{M.}~\bibnamefont{Abkarian}},
  \bibinfo{author}{\bibfnamefont{M.}~\bibnamefont{Faivre}}, \bibnamefont{and}
  \bibinfo{author}{\bibfnamefont{A.}~\bibnamefont{Viallat}},
  \bibinfo{journal}{Phys. Rev. Lett.} \textbf{\bibinfo{volume}{98}},
  \bibinfo{pages}{188302} (\bibinfo{year}{2007}).

\bibitem[{\citenamefont{Skotheim and Secomb}(2007)}]{skotheim_red_2007}
\bibinfo{author}{\bibfnamefont{J.~M.} \bibnamefont{Skotheim}} \bibnamefont{and}
  \bibinfo{author}{\bibfnamefont{T.~W.} \bibnamefont{Secomb}},
  \bibinfo{journal}{Phys. Rev. Lett.} \textbf{\bibinfo{volume}{98}},
  \bibinfo{pages}{78301} (\bibinfo{year}{2007}).

\bibitem[{\citenamefont{Pfafferott et~al.}(1985)\citenamefont{Pfafferott, Nash,
  and Meiselman}}]{pfafferott_red_1985}
\bibinfo{author}{\bibfnamefont{C.}~\bibnamefont{Pfafferott}},
  \bibinfo{author}{\bibfnamefont{G.~B.} \bibnamefont{Nash}}, \bibnamefont{and}
  \bibinfo{author}{\bibfnamefont{H.~J.} \bibnamefont{Meiselman}},
  \bibinfo{journal}{Biophys. J.} \textbf{\bibinfo{volume}{47}},
  \bibinfo{pages}{695} (\bibinfo{year}{1985}).

\bibitem[{\citenamefont{Chien}(1987)}]{chien_red_1987}
\bibinfo{author}{\bibfnamefont{S.}~\bibnamefont{Chien}}, \bibinfo{journal}{Ann.
  Rev. Physiol.} \textbf{\bibinfo{volume}{49}}, \bibinfo{pages}{177}
  (\bibinfo{year}{1987}).

\bibitem[{\citenamefont{Forsyth et~al.}(2011)\citenamefont{Forsyth, Wan,
  Owrutsky, Abkarian, and Stone}}]{forsyth_multiscale_2011}
\bibinfo{author}{\bibfnamefont{A.~M.} \bibnamefont{Forsyth}},
  \bibinfo{author}{\bibfnamefont{J.}~\bibnamefont{Wan}},
  \bibinfo{author}{\bibfnamefont{P.~D.} \bibnamefont{Owrutsky}},
  \bibinfo{author}{\bibfnamefont{M.}~\bibnamefont{Abkarian}}, \bibnamefont{and}
  \bibinfo{author}{\bibfnamefont{H.~A.} \bibnamefont{Stone}},
  \bibinfo{journal}{Proc. Natl. Acad. Sci. {USA}}
  \textbf{\bibinfo{volume}{108}}, \bibinfo{pages}{10986}
  (\bibinfo{year}{2011}).

\bibitem[{\citenamefont{Chien et~al.}(1966)\citenamefont{Chien, Usami, Taylor,
  Lundberg, and Gregersen}}]{chien_effects_1966}
\bibinfo{author}{\bibfnamefont{S.}~\bibnamefont{Chien}},
  \bibinfo{author}{\bibfnamefont{S.}~\bibnamefont{Usami}},
  \bibinfo{author}{\bibfnamefont{H.~M.} \bibnamefont{Taylor}},
  \bibinfo{author}{\bibfnamefont{J.~L.} \bibnamefont{Lundberg}},
  \bibnamefont{and} \bibinfo{author}{\bibfnamefont{M.~I.}
  \bibnamefont{Gregersen}}, \bibinfo{journal}{J. Appl. Physiol.}
  \textbf{\bibinfo{volume}{21}}, \bibinfo{pages}{81} (\bibinfo{year}{1966}).

\bibitem[{\citenamefont{Picart et~al.}(1998)\citenamefont{Picart, Piau,
  Galliard, and Carpentier}}]{picart_human_1998}
\bibinfo{author}{\bibfnamefont{C.}~\bibnamefont{Picart}},
  \bibinfo{author}{\bibfnamefont{J.-M.} \bibnamefont{Piau}},
  \bibinfo{author}{\bibfnamefont{H.}~\bibnamefont{Galliard}}, \bibnamefont{and}
  \bibinfo{author}{\bibfnamefont{P.}~\bibnamefont{Carpentier}},
  \bibinfo{journal}{J. Rheol.} \textbf{\bibinfo{volume}{42}},
  \bibinfo{pages}{1} (\bibinfo{year}{1998}).

\bibitem[{\citenamefont{Nott and Brady}(1994)}]{nott_pressure-driven_1994}
\bibinfo{author}{\bibfnamefont{P.~R.} \bibnamefont{Nott}} \bibnamefont{and}
  \bibinfo{author}{\bibfnamefont{J.~F.} \bibnamefont{Brady}},
  \bibinfo{journal}{J. Fluid Mech.} \textbf{\bibinfo{volume}{275}},
  \bibinfo{pages}{157} (\bibinfo{year}{1994}).

\bibitem[{\citenamefont{Morris and Boulay}(1999)}]{morris_curvilinear_1999}
\bibinfo{author}{\bibfnamefont{J.~F.} \bibnamefont{Morris}} \bibnamefont{and}
  \bibinfo{author}{\bibfnamefont{F.}~\bibnamefont{Boulay}},
  \bibinfo{journal}{J. Rheol.} \textbf{\bibinfo{volume}{43}},
  \bibinfo{pages}{1213} (\bibinfo{year}{1999}).

\bibitem[{\citenamefont{Yurkovetsky and
  Morris}(2008)}]{yurkovetsky_particle_2008}
\bibinfo{author}{\bibfnamefont{Y.}~\bibnamefont{Yurkovetsky}} \bibnamefont{and}
  \bibinfo{author}{\bibfnamefont{J.~F.} \bibnamefont{Morris}},
  \bibinfo{journal}{J. Rheol.} \textbf{\bibinfo{volume}{52}},
  \bibinfo{pages}{141} (\bibinfo{year}{2008}).

\bibitem[{\citenamefont{Deboeuf et~al.}(2009)\citenamefont{Deboeuf, Gauthier,
  Martin, Yurkovetsky, and Morris}}]{deboeuf_particle_2009}
\bibinfo{author}{\bibfnamefont{A.}~\bibnamefont{Deboeuf}},
  \bibinfo{author}{\bibfnamefont{G.}~\bibnamefont{Gauthier}},
  \bibinfo{author}{\bibfnamefont{J.}~\bibnamefont{Martin}},
  \bibinfo{author}{\bibfnamefont{Y.}~\bibnamefont{Yurkovetsky}},
  \bibnamefont{and} \bibinfo{author}{\bibfnamefont{J.~F.}
  \bibnamefont{Morris}}, \bibinfo{journal}{Phys. Rev. Lett.}
  \textbf{\bibinfo{volume}{102}}, \bibinfo{pages}{108301}
  (\bibinfo{year}{2009}).

\bibitem[{\citenamefont{Ramachandran and
  Leighton}(2008)}]{ramachandran_influence_2008}
\bibinfo{author}{\bibfnamefont{A.}~\bibnamefont{Ramachandran}}
  \bibnamefont{and} \bibinfo{author}{\bibfnamefont{D.~T.}
  \bibnamefont{Leighton}}, \bibinfo{journal}{J. Fluid Mech.}
  \textbf{\bibinfo{volume}{603}}, \bibinfo{pages}{207} (\bibinfo{year}{2008}).

\bibitem[{\citenamefont{Malkin et~al.}(2006)\citenamefont{Malkin, Malkin, and
  Isayev}}]{malkin_rheology:_2006}
\bibinfo{author}{\bibfnamefont{A.~I.} \bibnamefont{Malkin}},
  \bibinfo{author}{\bibfnamefont{A.~Y.} \bibnamefont{Malkin}},
  \bibnamefont{and} \bibinfo{author}{\bibfnamefont{A.~I.}
  \bibnamefont{Isayev}}, \emph{\bibinfo{title}{Rheology: Concepts, Methods, And
  Applications}} (\bibinfo{publisher}{{ChemTec} Publishing},
  \bibinfo{year}{2006}).

\bibitem[{\citenamefont{Peskin}(2002)}]{peskin_immersed_2002}
\bibinfo{author}{\bibfnamefont{C.~S.} \bibnamefont{Peskin}},
  \bibinfo{journal}{Acta Numerica} \textbf{\bibinfo{volume}{11}},
  \bibinfo{pages}{479} (\bibinfo{year}{2002}).

\bibitem[{\citenamefont{Kr\"uger et~al.}(2011)\citenamefont{Kr\"uger, Varnik,
  and Raabe}}]{kruger_efficient_2011}
\bibinfo{author}{\bibfnamefont{T.}~\bibnamefont{Kr\"uger}},
  \bibinfo{author}{\bibfnamefont{F.}~\bibnamefont{Varnik}}, \bibnamefont{and}
  \bibinfo{author}{\bibfnamefont{D.}~\bibnamefont{Raabe}},
  \bibinfo{journal}{Comput. Math. Appl.} \textbf{\bibinfo{volume}{61}},
  \bibinfo{pages}{3485} (\bibinfo{year}{2011}).

\bibitem[{\citenamefont{Kr\"uger et~al.}(2013)\citenamefont{Kr\"uger, Gross,
  Raabe, and Varnik}}]{kruger_crossover_2013}
\bibinfo{author}{\bibfnamefont{T.}~\bibnamefont{Kr\"uger}},
  \bibinfo{author}{\bibfnamefont{M.}~\bibnamefont{Gross}},
  \bibinfo{author}{\bibfnamefont{D.}~\bibnamefont{Raabe}}, \bibnamefont{and}
  \bibinfo{author}{\bibfnamefont{F.}~\bibnamefont{Varnik}},
  \bibinfo{journal}{Soft Matter} \textbf{\bibinfo{volume}{9}},
  \bibinfo{pages}{9008} (\bibinfo{year}{2013}).

\bibitem[{\citenamefont{Skalak et~al.}(1973)\citenamefont{Skalak, Tozeren,
  Zarda, and Chien}}]{skalak_strain_1973}
\bibinfo{author}{\bibfnamefont{R.}~\bibnamefont{Skalak}},
  \bibinfo{author}{\bibfnamefont{A.}~\bibnamefont{Tozeren}},
  \bibinfo{author}{\bibfnamefont{R.~P.} \bibnamefont{Zarda}}, \bibnamefont{and}
  \bibinfo{author}{\bibfnamefont{S.}~\bibnamefont{Chien}},
  \bibinfo{journal}{Biophys. J.} \textbf{\bibinfo{volume}{13}},
  \bibinfo{pages}{245} (\bibinfo{year}{1973}).

\bibitem[{\citenamefont{Helfrich}(1973)}]{helfrich_elastic_1973}
\bibinfo{author}{\bibfnamefont{W.}~\bibnamefont{Helfrich}},
  \bibinfo{journal}{Z. Naturforsch. C} \textbf{\bibinfo{volume}{28}},
  \bibinfo{pages}{693} (\bibinfo{year}{1973}).

\bibitem[{\citenamefont{Fedosov et~al.}(2011)\citenamefont{Fedosov, Pan,
  Caswell, Gompper, and Karniadakis}}]{fedosov_predicting_2011}
\bibinfo{author}{\bibfnamefont{D.~A.} \bibnamefont{Fedosov}},
  \bibinfo{author}{\bibfnamefont{W.}~\bibnamefont{Pan}},
  \bibinfo{author}{\bibfnamefont{B.}~\bibnamefont{Caswell}},
  \bibinfo{author}{\bibfnamefont{G.}~\bibnamefont{Gompper}}, \bibnamefont{and}
  \bibinfo{author}{\bibfnamefont{G.~E.} \bibnamefont{Karniadakis}},
  \bibinfo{journal}{Proc. Natl. Acad. Sci. {USA}}
  \textbf{\bibinfo{volume}{108}}, \bibinfo{pages}{11772}
  (\bibinfo{year}{2011}).

\bibitem[{\citenamefont{Evans and Skalak}(1980)}]{evans_mechanics_1980}
\bibinfo{author}{\bibfnamefont{E.~A.} \bibnamefont{Evans}} \bibnamefont{and}
  \bibinfo{author}{\bibfnamefont{R.}~\bibnamefont{Skalak}},
  \emph{\bibinfo{title}{Mechanics and thermodynamics of biomembranes}}
  (\bibinfo{publisher}{{CRC}}, \bibinfo{year}{1980}).

\bibitem[{\citenamefont{Seifert}(1997)}]{seifert_configurations_1997}
\bibinfo{author}{\bibfnamefont{U.}~\bibnamefont{Seifert}},
  \bibinfo{journal}{Adv. Phys.} \textbf{\bibinfo{volume}{46}},
  \bibinfo{pages}{13} (\bibinfo{year}{1997}).

\bibitem[{\citenamefont{Peskin and Printz}(1993)}]{peskin_improved_1993}
\bibinfo{author}{\bibfnamefont{C.~S.} \bibnamefont{Peskin}} \bibnamefont{and}
  \bibinfo{author}{\bibfnamefont{B.~F.} \bibnamefont{Printz}},
  \bibinfo{journal}{J. Comput. Phys.} \textbf{\bibinfo{volume}{105}},
  \bibinfo{pages}{33} (\bibinfo{year}{1993}).

\bibitem[{\citenamefont{Svetina et~al.}(2004)\citenamefont{Svetina, Kuzman,
  Waugh, Ziherl, and Zeks}}]{svetina_cooperative_2004}
\bibinfo{author}{\bibfnamefont{S.}~\bibnamefont{Svetina}},
  \bibinfo{author}{\bibfnamefont{D.}~\bibnamefont{Kuzman}},
  \bibinfo{author}{\bibfnamefont{R.~E.} \bibnamefont{Waugh}},
  \bibinfo{author}{\bibfnamefont{P.}~\bibnamefont{Ziherl}}, \bibnamefont{and}
  \bibinfo{author}{\bibfnamefont{B.}~\bibnamefont{Zeks}},
  \bibinfo{journal}{Bioelectroch.} \textbf{\bibinfo{volume}{62}},
  \bibinfo{pages}{107} (\bibinfo{year}{2004}).

\bibitem[{\citenamefont{Gompper and Schick}(2008)}]{gompper_soft_2008}
\bibinfo{author}{\bibfnamefont{G.}~\bibnamefont{Gompper}} \bibnamefont{and}
  \bibinfo{author}{\bibfnamefont{M.}~\bibnamefont{Schick}},
  \emph{\bibinfo{title}{Soft Matter: Lipid Bilayers and Red Blood Cells}}
  (\bibinfo{publisher}{Wiley-{VCH}}, \bibinfo{year}{2008}).

\bibitem[{\citenamefont{Feng and Michaelides}(2004)}]{feng_immersed_2004}
\bibinfo{author}{\bibfnamefont{Z.-G.} \bibnamefont{Feng}} \bibnamefont{and}
  \bibinfo{author}{\bibfnamefont{E.~E.} \bibnamefont{Michaelides}},
  \bibinfo{journal}{J. Comput. Phys.} \textbf{\bibinfo{volume}{195}},
  \bibinfo{pages}{602} (\bibinfo{year}{2004}).

\bibitem[{\citenamefont{Succi}(2001)}]{succi_lattice_2001}
\bibinfo{author}{\bibfnamefont{S.}~\bibnamefont{Succi}},
  \emph{\bibinfo{title}{The Lattice Boltzmann Equation for Fluid Dynamics and
  Beyond}} (\bibinfo{publisher}{Oxford University Press},
  \bibinfo{year}{2001}).

\bibitem[{\citenamefont{Ladd}(1994{\natexlab{a}})}]{ladd_numerical_1994}
\bibinfo{author}{\bibfnamefont{A.~J.~C.} \bibnamefont{Ladd}},
  \bibinfo{journal}{J. Fluid Mech.} \textbf{\bibinfo{volume}{271}},
  \bibinfo{pages}{285} (\bibinfo{year}{1994}{\natexlab{a}}).

\bibitem[{\citenamefont{Ladd}(1994{\natexlab{b}})}]{ladd_numerical_1994-1}
\bibinfo{author}{\bibfnamefont{A.~J.~C.} \bibnamefont{Ladd}},
  \bibinfo{journal}{J. Fluid Mech.} \textbf{\bibinfo{volume}{271}},
  \bibinfo{pages}{311} (\bibinfo{year}{1994}{\natexlab{b}}).

\bibitem[{\citenamefont{Ladd and Verberg}(2001)}]{ladd_lattice-boltzmann_2001}
\bibinfo{author}{\bibfnamefont{A.~J.~C.} \bibnamefont{Ladd}} \bibnamefont{and}
  \bibinfo{author}{\bibfnamefont{R.}~\bibnamefont{Verberg}},
  \bibinfo{journal}{J. Stat. Phys.} \textbf{\bibinfo{volume}{104}},
  \bibinfo{pages}{1191} (\bibinfo{year}{2001}).

\bibitem[{\citenamefont{Batchelor}(1970)}]{batchelor_stress_1970}
\bibinfo{author}{\bibfnamefont{G.~K.} \bibnamefont{Batchelor}},
  \bibinfo{journal}{J. Fluid Mech.} \textbf{\bibinfo{volume}{41}},
  \bibinfo{pages}{545} (\bibinfo{year}{1970}).

\bibitem[{\citenamefont{Pozrikidis}(1995)}]{pozrikidis_finite_1995}
\bibinfo{author}{\bibfnamefont{C.}~\bibnamefont{Pozrikidis}},
  \bibinfo{journal}{J. Fluid Mech.} \textbf{\bibinfo{volume}{297}},
  \bibinfo{pages}{123} (\bibinfo{year}{1995}).

\bibitem[{\citenamefont{Bagchi and Kalluri}(2010)}]{bagchi_rheology_2010}
\bibinfo{author}{\bibfnamefont{P.}~\bibnamefont{Bagchi}} \bibnamefont{and}
  \bibinfo{author}{\bibfnamefont{R.~M.} \bibnamefont{Kalluri}},
  \bibinfo{journal}{Phys. Rev. E} \textbf{\bibinfo{volume}{81}},
  \bibinfo{pages}{056320} (\bibinfo{year}{2010}).

\bibitem[{\citenamefont{Mandal et~al.}(2012)\citenamefont{Mandal, Gross, Raabe,
  and Varnik}}]{mandal_heterogeneous_2012}
\bibinfo{author}{\bibfnamefont{S.}~\bibnamefont{Mandal}},
  \bibinfo{author}{\bibfnamefont{M.}~\bibnamefont{Gross}},
  \bibinfo{author}{\bibfnamefont{D.}~\bibnamefont{Raabe}}, \bibnamefont{and}
  \bibinfo{author}{\bibfnamefont{F.}~\bibnamefont{Varnik}},
  \bibinfo{journal}{Phys. Rev. Lett.} \textbf{\bibinfo{volume}{108}},
  \bibinfo{pages}{098301} (\bibinfo{year}{2012}).

\bibitem[{\citenamefont{Ackerson and
  Pusey}(1988)}]{ackerson_shear-induced_1988}
\bibinfo{author}{\bibfnamefont{B.~J.} \bibnamefont{Ackerson}} \bibnamefont{and}
  \bibinfo{author}{\bibfnamefont{P.~N.} \bibnamefont{Pusey}},
  \bibinfo{journal}{Phys. Rev. Lett.} \textbf{\bibinfo{volume}{61}},
  \bibinfo{pages}{1033} (\bibinfo{year}{1988}).

\bibitem[{\citenamefont{Ackerson}(1990)}]{ackerson_shear_1990}
\bibinfo{author}{\bibfnamefont{B.~J.} \bibnamefont{Ackerson}},
  \bibinfo{journal}{J. Rheol.} \textbf{\bibinfo{volume}{34}},
  \bibinfo{pages}{553} (\bibinfo{year}{1990}).

\bibitem[{\citenamefont{Cheng et~al.}(2012)\citenamefont{Cheng, Xu, Rice,
  Dinner, and Cohen}}]{cheng_assembly_2012}
\bibinfo{author}{\bibfnamefont{X.}~\bibnamefont{Cheng}},
  \bibinfo{author}{\bibfnamefont{X.}~\bibnamefont{Xu}},
  \bibinfo{author}{\bibfnamefont{S.~A.} \bibnamefont{Rice}},
  \bibinfo{author}{\bibfnamefont{A.~R.} \bibnamefont{Dinner}},
  \bibnamefont{and} \bibinfo{author}{\bibfnamefont{I.}~\bibnamefont{Cohen}},
  \bibinfo{journal}{Proc. Natl. Acad. Sci. {USA}}
  \textbf{\bibinfo{volume}{109}}, \bibinfo{pages}{63} (\bibinfo{year}{2012}).

\bibitem[{\citenamefont{Zurita-Gotor et~al.}(2012)\citenamefont{Zurita-Gotor,
  Bławzdziewicz, and Wajnryb}}]{zurita-gotor_layering_2012}
\bibinfo{author}{\bibfnamefont{M.}~\bibnamefont{Zurita-Gotor}},
  \bibinfo{author}{\bibfnamefont{J.}~\bibnamefont{Bławzdziewicz}},
  \bibnamefont{and} \bibinfo{author}{\bibfnamefont{E.}~\bibnamefont{Wajnryb}},
  \bibinfo{journal}{Phys. Rev. Lett.} \textbf{\bibinfo{volume}{108}},
  \bibinfo{pages}{068301} (\bibinfo{year}{2012}).

\bibitem[{\citenamefont{Einstein}(1906)}]{einstein_new_1906}
\bibinfo{author}{\bibfnamefont{A.}~\bibnamefont{Einstein}},
  \bibinfo{journal}{Ann. Phys.} \textbf{\bibinfo{volume}{19}},
  \bibinfo{pages}{289} (\bibinfo{year}{1906}).

\bibitem[{\citenamefont{Jeffery}(1922)}]{jeffery_motion_1922}
\bibinfo{author}{\bibfnamefont{G.~B.} \bibnamefont{Jeffery}},
  \bibinfo{journal}{Proc. Roy. Soc. Lond. A Mat.}
  \textbf{\bibinfo{volume}{102}}, \bibinfo{pages}{161} (\bibinfo{year}{1922}).

\bibitem[{\citenamefont{Yazdani and Bagchi}(2013)}]{yazdani_influence_2013}
\bibinfo{author}{\bibfnamefont{A.}~\bibnamefont{Yazdani}} \bibnamefont{and}
  \bibinfo{author}{\bibfnamefont{P.}~\bibnamefont{Bagchi}},
  \bibinfo{journal}{J. Fluid Mech.} \textbf{\bibinfo{volume}{718}},
  \bibinfo{pages}{569} (\bibinfo{year}{2013}).

\bibitem[{\citenamefont{Bagchi and Kalluri}(2011)}]{bagchi_dynamic_2011}
\bibinfo{author}{\bibfnamefont{P.}~\bibnamefont{Bagchi}} \bibnamefont{and}
  \bibinfo{author}{\bibfnamefont{R.~M.} \bibnamefont{Kalluri}},
  \bibinfo{journal}{J. Fluid Mech.} \textbf{\bibinfo{volume}{669}},
  \bibinfo{pages}{498} (\bibinfo{year}{2011}).

\bibitem[{\citenamefont{Gao et~al.}(2012)\citenamefont{Gao, Hu, and
  Castañeda}}]{gao_shape_2012}
\bibinfo{author}{\bibfnamefont{T.}~\bibnamefont{Gao}},
  \bibinfo{author}{\bibfnamefont{H.~H.} \bibnamefont{Hu}}, \bibnamefont{and}
  \bibinfo{author}{\bibfnamefont{P.~P.} \bibnamefont{Castañeda}},
  \bibinfo{journal}{Phys. Rev. Lett.} \textbf{\bibinfo{volume}{108}},
  \bibinfo{pages}{058302} (\bibinfo{year}{2012}).

\bibitem[{\citenamefont{Goddard and Miller}(1967)}]{goddard_nonlinear_1967}
\bibinfo{author}{\bibfnamefont{J.~D.} \bibnamefont{Goddard}} \bibnamefont{and}
  \bibinfo{author}{\bibfnamefont{C.}~\bibnamefont{Miller}},
  \bibinfo{journal}{J. Fluid Mech.} \textbf{\bibinfo{volume}{28}},
  \bibinfo{pages}{657} (\bibinfo{year}{1967}).

\bibitem[{\citenamefont{Roscoe}(1967)}]{roscoe_rheology_1967}
\bibinfo{author}{\bibfnamefont{R.}~\bibnamefont{Roscoe}}, \bibinfo{journal}{J.
  Fluid Mech.} \textbf{\bibinfo{volume}{28}}, \bibinfo{pages}{273}
  (\bibinfo{year}{1967}).

\bibitem[{\citenamefont{Barth\`{e}s-Biesel and
  Chhim}(1981)}]{barthes-biesel_constitutive_1981}
\bibinfo{author}{\bibfnamefont{D.}~\bibnamefont{Barth\`{e}s-Biesel}}
  \bibnamefont{and} \bibinfo{author}{\bibfnamefont{V.}~\bibnamefont{Chhim}},
  \bibinfo{journal}{Int. J. Multiphase Flow} \textbf{\bibinfo{volume}{7}},
  \bibinfo{pages}{493} (\bibinfo{year}{1981}).

\bibitem[{\citenamefont{Navot}(1998)}]{navot_elastic_1998}
\bibinfo{author}{\bibfnamefont{Y.}~\bibnamefont{Navot}},
  \bibinfo{journal}{Phys. Fluids} \textbf{\bibinfo{volume}{10}},
  \bibinfo{pages}{1819} (\bibinfo{year}{1998}).

\bibitem[{\citenamefont{Barth\`{e}s-Biesel}(1980)}]{barthes-biesel_motion_1980}
\bibinfo{author}{\bibfnamefont{D.}~\bibnamefont{Barth\`{e}s-Biesel}},
  \bibinfo{journal}{J. Fluid Mech.} \textbf{\bibinfo{volume}{100}},
  \bibinfo{pages}{831} (\bibinfo{year}{1980}).

\bibitem[{\citenamefont{Guazzelli and Morris}(2012)}]{guazzelli_physical_2012}
\bibinfo{author}{\bibfnamefont{E.}~\bibnamefont{Guazzelli}} \bibnamefont{and}
  \bibinfo{author}{\bibfnamefont{J.~F.} \bibnamefont{Morris}},
  \emph{\bibinfo{title}{A physical introduction to suspension dynamics}}
  (\bibinfo{publisher}{Cambridge University Press},
  \bibinfo{address}{Cambridge; New York}, \bibinfo{year}{2012}).

\bibitem[{\citenamefont{Donev et~al.}(2004)\citenamefont{Donev, Cisse, Sachs,
  Variano, Stillinger, Connelly, Torquato, and Chaikin}}]{donev_improving_2004}
\bibinfo{author}{\bibfnamefont{A.}~\bibnamefont{Donev}},
  \bibinfo{author}{\bibfnamefont{I.}~\bibnamefont{Cisse}},
  \bibinfo{author}{\bibfnamefont{D.}~\bibnamefont{Sachs}},
  \bibinfo{author}{\bibfnamefont{E.~A.} \bibnamefont{Variano}},
  \bibinfo{author}{\bibfnamefont{F.~H.} \bibnamefont{Stillinger}},
  \bibinfo{author}{\bibfnamefont{R.}~\bibnamefont{Connelly}},
  \bibinfo{author}{\bibfnamefont{S.}~\bibnamefont{Torquato}}, \bibnamefont{and}
  \bibinfo{author}{\bibfnamefont{P.~M.} \bibnamefont{Chaikin}},
  \bibinfo{journal}{Science} \textbf{\bibinfo{volume}{303}},
  \bibinfo{pages}{990} (\bibinfo{year}{2004}).

\bibitem[{\citenamefont{Bolton and Weaire}(1990)}]{bolton_rigidity_1990}
\bibinfo{author}{\bibfnamefont{F.}~\bibnamefont{Bolton}} \bibnamefont{and}
  \bibinfo{author}{\bibfnamefont{D.}~\bibnamefont{Weaire}},
  \bibinfo{journal}{Phys. Rev. Lett.} \textbf{\bibinfo{volume}{65}},
  \bibinfo{pages}{3449} (\bibinfo{year}{1990}).

\bibitem[{\citenamefont{Mason et~al.}(1996)\citenamefont{Mason, Bibette, and
  Weitz}}]{mason_yielding_1996}
\bibinfo{author}{\bibfnamefont{T.}~\bibnamefont{Mason}},
  \bibinfo{author}{\bibfnamefont{J.}~\bibnamefont{Bibette}}, \bibnamefont{and}
  \bibinfo{author}{\bibfnamefont{D.}~\bibnamefont{Weitz}}, \bibinfo{journal}{J.
  Coll. Int. Sci.} \textbf{\bibinfo{volume}{179}}, \bibinfo{pages}{439}
  (\bibinfo{year}{1996}).

\bibitem[{\citenamefont{van Hecke}(2010)}]{van_hecke_jamming_2010}
\bibinfo{author}{\bibfnamefont{M.}~\bibnamefont{van Hecke}},
  \bibinfo{journal}{J. Phys.: Cond. Mat.} \textbf{\bibinfo{volume}{22}},
  \bibinfo{pages}{033101} (\bibinfo{year}{2010}).

\bibitem[{\citenamefont{Landau and Lifshitz}(1986)}]{landau_theory_1986}
\bibinfo{author}{\bibfnamefont{L.~D.} \bibnamefont{Landau}} \bibnamefont{and}
  \bibinfo{author}{\bibfnamefont{E.~M.} \bibnamefont{Lifshitz}},
  \emph{\bibinfo{title}{Theory of Elasticity}}
  (\bibinfo{publisher}{Butterworth-Heinemann}, \bibinfo{address}{Oxford},
  \bibinfo{year}{1986}).

\bibitem[{\citenamefont{Siber and Ziherl}(2013)}]{siber_many-body_2013}
\bibinfo{author}{\bibfnamefont{A.}~\bibnamefont{Siber}} \bibnamefont{and}
  \bibinfo{author}{\bibfnamefont{P.}~\bibnamefont{Ziherl}},
  \bibinfo{journal}{Phys. Rev. Lett.} \textbf{\bibinfo{volume}{110}},
  \bibinfo{pages}{214301} (\bibinfo{year}{2013}).

\bibitem[{\citenamefont{{O'Hern} et~al.}(2003)\citenamefont{{O'Hern}, Silbert,
  Liu, and Nagel}}]{ohern_jamming_2003}
\bibinfo{author}{\bibfnamefont{C.~S.} \bibnamefont{{O'Hern}}},
  \bibinfo{author}{\bibfnamefont{L.~E.} \bibnamefont{Silbert}},
  \bibinfo{author}{\bibfnamefont{A.~J.} \bibnamefont{Liu}}, \bibnamefont{and}
  \bibinfo{author}{\bibfnamefont{S.~R.} \bibnamefont{Nagel}},
  \bibinfo{journal}{Phys. Rev. E} \textbf{\bibinfo{volume}{68}},
  \bibinfo{pages}{011306} (\bibinfo{year}{2003}).

\bibitem[{\citenamefont{Zydney et~al.}(1991)\citenamefont{Zydney, Oliver, and
  Colton}}]{zydney_constitutive_1991}
\bibinfo{author}{\bibfnamefont{A.~L.} \bibnamefont{Zydney}},
  \bibinfo{author}{\bibfnamefont{J.~D.} \bibnamefont{Oliver}},
  \bibnamefont{and} \bibinfo{author}{\bibfnamefont{C.~K.}
  \bibnamefont{Colton}}, \bibinfo{journal}{J. Rheol.}
  \textbf{\bibinfo{volume}{35}}, \bibinfo{pages}{1639} (\bibinfo{year}{1991}).

\bibitem[{\citenamefont{Olsson and Teitel}(2007)}]{olsson_critical_2007}
\bibinfo{author}{\bibfnamefont{P.}~\bibnamefont{Olsson}} \bibnamefont{and}
  \bibinfo{author}{\bibfnamefont{S.}~\bibnamefont{Teitel}},
  \bibinfo{journal}{Phys. Rev. Lett.} \textbf{\bibinfo{volume}{99}},
  \bibinfo{pages}{178001} (\bibinfo{year}{2007}).

\bibitem[{\citenamefont{Sierou and Brady}(2002)}]{sierou_rheology_2002}
\bibinfo{author}{\bibfnamefont{A.}~\bibnamefont{Sierou}} \bibnamefont{and}
  \bibinfo{author}{\bibfnamefont{J.~F.} \bibnamefont{Brady}},
  \bibinfo{journal}{J. Rheol.} \textbf{\bibinfo{volume}{46}},
  \bibinfo{pages}{1031} (\bibinfo{year}{2002}).

\bibitem[{\citenamefont{Ramanujan and
  Pozrikidis}(1998)}]{ramanujan_deformation_1998}
\bibinfo{author}{\bibfnamefont{S.}~\bibnamefont{Ramanujan}} \bibnamefont{and}
  \bibinfo{author}{\bibfnamefont{C.}~\bibnamefont{Pozrikidis}},
  \bibinfo{journal}{J. Fluid Mech.} \textbf{\bibinfo{volume}{361}},
  \bibinfo{pages}{117} (\bibinfo{year}{1998}).

\bibitem[{\citenamefont{Zirnsak et~al.}(1994)\citenamefont{Zirnsak, Hur, and
  Boger}}]{zirnsak_normal_1994}
\bibinfo{author}{\bibfnamefont{M.}~\bibnamefont{Zirnsak}},
  \bibinfo{author}{\bibfnamefont{D.}~\bibnamefont{Hur}}, \bibnamefont{and}
  \bibinfo{author}{\bibfnamefont{D.}~\bibnamefont{Boger}}, \bibinfo{journal}{J.
  non-Newton. Fluid Mech.} \textbf{\bibinfo{volume}{54}}, \bibinfo{pages}{153}
  (\bibinfo{year}{1994}).

\bibitem[{\citenamefont{Bertevas et~al.}(2010)\citenamefont{Bertevas, Fan, and
  Tanner}}]{bertevas_simulation_2010}
\bibinfo{author}{\bibfnamefont{E.}~\bibnamefont{Bertevas}},
  \bibinfo{author}{\bibfnamefont{X.}~\bibnamefont{Fan}}, \bibnamefont{and}
  \bibinfo{author}{\bibfnamefont{R.~I.} \bibnamefont{Tanner}},
  \bibinfo{journal}{Rheol. Acta} \textbf{\bibinfo{volume}{49}},
  \bibinfo{pages}{53} (\bibinfo{year}{2010}).

\bibitem[{\citenamefont{Snabre and Mills}(1999)}]{snabre_rheology_1999}
\bibinfo{author}{\bibfnamefont{P.}~\bibnamefont{Snabre}} \bibnamefont{and}
  \bibinfo{author}{\bibfnamefont{P.}~\bibnamefont{Mills}},
  \bibinfo{journal}{Colloids Surf. A} \textbf{\bibinfo{volume}{152}},
  \bibinfo{pages}{79} (\bibinfo{year}{1999}).

\bibitem[{\citenamefont{Pal}(2003)}]{pal_rheology_2003}
\bibinfo{author}{\bibfnamefont{R.}~\bibnamefont{Pal}}, \bibinfo{journal}{J.
  Biomech.} \textbf{\bibinfo{volume}{36}}, \bibinfo{pages}{981}
  (\bibinfo{year}{2003}).

\bibitem[{\citenamefont{Krieger and Dougherty}(1959)}]{krieger_mechanism_1959}
\bibinfo{author}{\bibfnamefont{I.~M.} \bibnamefont{Krieger}} \bibnamefont{and}
  \bibinfo{author}{\bibfnamefont{T.~J.} \bibnamefont{Dougherty}},
  \bibinfo{journal}{Trans. Soc. Rheol.} \textbf{\bibinfo{volume}{3}},
  \bibinfo{pages}{137} (\bibinfo{year}{1959}).

\bibitem[{\citenamefont{Gross}(2014)}]{gross_inprep}
\bibinfo{author}{\bibfnamefont{M.}~\bibnamefont{Gross}},
  \bibinfo{journal}{unpublished}  (\bibinfo{year}{2014}).

\bibitem[{\citenamefont{Batchelor and
  Green}(1972)}]{batchelor_determination_1972}
\bibinfo{author}{\bibfnamefont{G.~K.} \bibnamefont{Batchelor}}
  \bibnamefont{and} \bibinfo{author}{\bibfnamefont{J.~T.} \bibnamefont{Green}},
  \bibinfo{journal}{J. Fluid Mech.} \textbf{\bibinfo{volume}{56}},
  \bibinfo{pages}{401} (\bibinfo{year}{1972}).

\bibitem[{\citenamefont{Guth and Simha}(1936)}]{guth_untersuchungen_1936}
\bibinfo{author}{\bibfnamefont{E.}~\bibnamefont{Guth}} \bibnamefont{and}
  \bibinfo{author}{\bibfnamefont{R.}~\bibnamefont{Simha}},
  \bibinfo{journal}{Kolloid-Zeitschrift} \textbf{\bibinfo{volume}{74}},
  \bibinfo{pages}{266} (\bibinfo{year}{1936}).

\bibitem[{\citenamefont{Davit and Peyla}(2008)}]{davit_intriguing_2008}
\bibinfo{author}{\bibfnamefont{Y.}~\bibnamefont{Davit}} \bibnamefont{and}
  \bibinfo{author}{\bibfnamefont{P.}~\bibnamefont{Peyla}},
  \bibinfo{journal}{{EPL} (Europhysics Lett.)} \textbf{\bibinfo{volume}{83}},
  \bibinfo{pages}{64001} (\bibinfo{year}{2008}).

\bibitem[{\citenamefont{Swan and Brady}(2010)}]{swan_particle_2010}
\bibinfo{author}{\bibfnamefont{J.~W.} \bibnamefont{Swan}} \bibnamefont{and}
  \bibinfo{author}{\bibfnamefont{J.~F.} \bibnamefont{Brady}},
  \bibinfo{journal}{Phys. Fluids} \textbf{\bibinfo{volume}{22}},
  \bibinfo{pages}{103301} (\bibinfo{year}{2010}).

\bibitem[{\citenamefont{Peyla and Verdier}(2011)}]{peyla_new_2011}
\bibinfo{author}{\bibfnamefont{P.}~\bibnamefont{Peyla}} \bibnamefont{and}
  \bibinfo{author}{\bibfnamefont{C.}~\bibnamefont{Verdier}},
  \bibinfo{journal}{{EPL} (Europhysics Lett.)} \textbf{\bibinfo{volume}{94}},
  \bibinfo{pages}{44001} (\bibinfo{year}{2011}).

\bibitem[{\citenamefont{Sangani et~al.}(2011)\citenamefont{Sangani, Acrivos,
  and Peyla}}]{sangani_roles_2011}
\bibinfo{author}{\bibfnamefont{A.~S.} \bibnamefont{Sangani}},
  \bibinfo{author}{\bibfnamefont{A.}~\bibnamefont{Acrivos}}, \bibnamefont{and}
  \bibinfo{author}{\bibfnamefont{P.}~\bibnamefont{Peyla}},
  \bibinfo{journal}{Phys. Fluids} \textbf{\bibinfo{volume}{23}},
  \bibinfo{pages}{083302} (\bibinfo{year}{2011}).

\bibitem[{\citenamefont{Kaoui et~al.}(2012)\citenamefont{Kaoui, Kr\"uger, and
  Harting}}]{kaoui_how_2012}
\bibinfo{author}{\bibfnamefont{B.}~\bibnamefont{Kaoui}},
  \bibinfo{author}{\bibfnamefont{T.}~\bibnamefont{Kr\"uger}}, \bibnamefont{and}
  \bibinfo{author}{\bibfnamefont{J.}~\bibnamefont{Harting}},
  \bibinfo{journal}{Soft Matter} \textbf{\bibinfo{volume}{8}},
  \bibinfo{pages}{9246} (\bibinfo{year}{2012}).

\bibitem[{\citenamefont{Fahraeus and
  Lindqvist}(1931)}]{faahraeus_viscosity_1931}
\bibinfo{author}{\bibfnamefont{R.}~\bibnamefont{Fahraeus}} \bibnamefont{and}
  \bibinfo{author}{\bibfnamefont{T.}~\bibnamefont{Lindqvist}},
  \bibinfo{journal}{Am. J. Physiol.} \textbf{\bibinfo{volume}{96}},
  \bibinfo{pages}{562} (\bibinfo{year}{1931}).

\bibitem[{\citenamefont{Levant et~al.}(2012)\citenamefont{Levant, Deschamps,
  Afik, and Steinberg}}]{levant_characteristic_2012}
\bibinfo{author}{\bibfnamefont{M.}~\bibnamefont{Levant}},
  \bibinfo{author}{\bibfnamefont{J.}~\bibnamefont{Deschamps}},
  \bibinfo{author}{\bibfnamefont{E.}~\bibnamefont{Afik}}, \bibnamefont{and}
  \bibinfo{author}{\bibfnamefont{V.}~\bibnamefont{Steinberg}},
  \bibinfo{journal}{Phys. Rev. E} \textbf{\bibinfo{volume}{85}},
  \bibinfo{pages}{056306} (\bibinfo{year}{2012}).

\bibitem[{\citenamefont{Langlois et~al.}(2008)\citenamefont{Langlois, Hutzler,
  and Weaire}}]{langlois_rheological_2008}
\bibinfo{author}{\bibfnamefont{V.~J.} \bibnamefont{Langlois}},
  \bibinfo{author}{\bibfnamefont{S.}~\bibnamefont{Hutzler}}, \bibnamefont{and}
  \bibinfo{author}{\bibfnamefont{D.}~\bibnamefont{Weaire}},
  \bibinfo{journal}{Phys. Rev. E} \textbf{\bibinfo{volume}{78}},
  \bibinfo{pages}{021401} (\bibinfo{year}{2008}).

\bibitem[{\citenamefont{Chaudhuri et~al.}(2012)\citenamefont{Chaudhuri,
  Berthier, and Bocquet}}]{chaudhuri_inhomogeneous_2012}
\bibinfo{author}{\bibfnamefont{P.}~\bibnamefont{Chaudhuri}},
  \bibinfo{author}{\bibfnamefont{L.}~\bibnamefont{Berthier}}, \bibnamefont{and}
  \bibinfo{author}{\bibfnamefont{L.}~\bibnamefont{Bocquet}},
  \bibinfo{journal}{Phys. Rev. E} \textbf{\bibinfo{volume}{85}},
  \bibinfo{pages}{021503} (\bibinfo{year}{2012}).

\bibitem[{\citenamefont{Mansard et~al.}(2013)\citenamefont{Mansard, Colin,
  Chaudhuri, and Bocquet}}]{mansard_molecular_2013}
\bibinfo{author}{\bibfnamefont{V.}~\bibnamefont{Mansard}},
  \bibinfo{author}{\bibfnamefont{A.}~\bibnamefont{Colin}},
  \bibinfo{author}{\bibfnamefont{P.}~\bibnamefont{Chaudhuri}},
  \bibnamefont{and} \bibinfo{author}{\bibfnamefont{L.}~\bibnamefont{Bocquet}},
  \bibinfo{journal}{Soft Matter} \textbf{\bibinfo{volume}{9}},
  \bibinfo{pages}{7489} (\bibinfo{year}{2013}).

\bibitem[{\citenamefont{Bocquet et~al.}(2009)\citenamefont{Bocquet, Colin, and
  Ajdari}}]{bocquet_kinetic_2009}
\bibinfo{author}{\bibfnamefont{L.}~\bibnamefont{Bocquet}},
  \bibinfo{author}{\bibfnamefont{A.}~\bibnamefont{Colin}}, \bibnamefont{and}
  \bibinfo{author}{\bibfnamefont{A.}~\bibnamefont{Ajdari}},
  \bibinfo{journal}{Phys. Rev. Lett.} \textbf{\bibinfo{volume}{103}},
  \bibinfo{pages}{036001} (\bibinfo{year}{2009}).

\bibitem[{\citenamefont{Denkov et~al.}(2008)\citenamefont{Denkov, Tcholakova,
  Golemanov, Ananthapadmanabhan, and Lips}}]{denkov_viscous_2008}
\bibinfo{author}{\bibfnamefont{N.~D.} \bibnamefont{Denkov}},
  \bibinfo{author}{\bibfnamefont{S.}~\bibnamefont{Tcholakova}},
  \bibinfo{author}{\bibfnamefont{K.}~\bibnamefont{Golemanov}},
  \bibinfo{author}{\bibfnamefont{K.~P.} \bibnamefont{Ananthapadmanabhan}},
  \bibnamefont{and} \bibinfo{author}{\bibfnamefont{A.}~\bibnamefont{Lips}},
  \bibinfo{journal}{Phys. Rev. Lett.} \textbf{\bibinfo{volume}{100}},
  \bibinfo{pages}{138301} (\bibinfo{year}{2008}).

\bibitem[{\citenamefont{Seth et~al.}(2011)\citenamefont{Seth, Mohan,
  Locatelli-Champagne, Cloitre, and Bonnecaze}}]{seth_micromechanical_2011}
\bibinfo{author}{\bibfnamefont{J.~R.} \bibnamefont{Seth}},
  \bibinfo{author}{\bibfnamefont{L.}~\bibnamefont{Mohan}},
  \bibinfo{author}{\bibfnamefont{C.}~\bibnamefont{Locatelli-Champagne}},
  \bibinfo{author}{\bibfnamefont{M.}~\bibnamefont{Cloitre}}, \bibnamefont{and}
  \bibinfo{author}{\bibfnamefont{R.~T.} \bibnamefont{Bonnecaze}},
  \bibinfo{journal}{Nat. Mater.} \textbf{\bibinfo{volume}{10}},
  \bibinfo{pages}{838} (\bibinfo{year}{2011}).

\bibitem[{\citenamefont{Schmitt et~al.}(1995)\citenamefont{Schmitt, Marques,
  and Lequeux}}]{schmitt_shear-induced_1995}
\bibinfo{author}{\bibfnamefont{V.}~\bibnamefont{Schmitt}},
  \bibinfo{author}{\bibfnamefont{C.~M.} \bibnamefont{Marques}},
  \bibnamefont{and} \bibinfo{author}{\bibfnamefont{F.}~\bibnamefont{Lequeux}},
  \bibinfo{journal}{Phys. Rev. E} \textbf{\bibinfo{volume}{52}},
  \bibinfo{pages}{4009} (\bibinfo{year}{1995}).

\bibitem[{\citenamefont{Besseling et~al.}(2010)\citenamefont{Besseling, Isa,
  Ballesta, Petekidis, Cates, and Poon}}]{besseling_shear_2010}
\bibinfo{author}{\bibfnamefont{R.}~\bibnamefont{Besseling}},
  \bibinfo{author}{\bibfnamefont{L.}~\bibnamefont{Isa}},
  \bibinfo{author}{\bibfnamefont{P.}~\bibnamefont{Ballesta}},
  \bibinfo{author}{\bibfnamefont{G.}~\bibnamefont{Petekidis}},
  \bibinfo{author}{\bibfnamefont{M.~E.} \bibnamefont{Cates}}, \bibnamefont{and}
  \bibinfo{author}{\bibfnamefont{W.~C.~K.} \bibnamefont{Poon}},
  \bibinfo{journal}{Phys. Rev. Lett.} \textbf{\bibinfo{volume}{105}},
  \bibinfo{pages}{268301} (\bibinfo{year}{2010}).

\bibitem[{\citenamefont{Vlahovska et~al.}(2009)\citenamefont{Vlahovska,
  Podgorski, and Misbah}}]{vlahovska_vesicles_2009}
\bibinfo{author}{\bibfnamefont{P.~M.} \bibnamefont{Vlahovska}},
  \bibinfo{author}{\bibfnamefont{T.}~\bibnamefont{Podgorski}},
  \bibnamefont{and} \bibinfo{author}{\bibfnamefont{C.}~\bibnamefont{Misbah}},
  \bibinfo{journal}{Compt. Rend. Phys.} \textbf{\bibinfo{volume}{10}},
  \bibinfo{pages}{775} (\bibinfo{year}{2009}).

\end{thebibliography}

\end{document}